\documentclass[aps,prstab,twocolumn,groupedaddress,nofootinbib,floatfix]
{revtex4-1}

\usepackage{amsmath}
\usepackage{graphicx}
\usepackage[caption=false]{subfig}
\usepackage{color,soul}

\usepackage{relsize}
\usepackage{xspace}

\newcommand{\unit}[1]{\ensuremath{\,\mathrm{#1}}}

\begin{document}


\title{
Simulation of the transit-time optical stochastic cooling process in the Cornell Electron Storage Ring
}



\author{S. T. Wang}
\email[]{sw565@cornell.edu}
\author{M. B. Andorf}
\author{I. V. Bazarov}
\author{W. F. Bergan}
\author{V. Khachatryan}
\author{J. M. Maxson}
\author{D. L. Rubin}
\affiliation{Cornell Laboratory for Accelerator-based Sciences and 
Education, Cornell University, Ithaca, NY 14853}




\begin{abstract}
In preparation for a demonstration of optical stochastic cooling in the Cornell Electron Storage Ring (CESR) we have developed a particle tracking simulation to study the relevant beam dynamics. Optical radiation emitted in the pickup undulator gives a momentum kick to that same particle in the kicker undulator. The optics of the electron bypass from pickup to kicker couples betatron amplitude and momentum offset to path length so that the momentum kick reduces emittance and momentum spread. Nearby electrons contribute an incoherent noise. Layout of the bypass line is presented that accommodates optics with a range of transverse and longitudinal cooling parameters. The simulation is used to determine cooling rates and their dependence on bunch and lattice parameters for bypass optics with distinct emittance and momentum acceptance.
\end{abstract}


\maketitle

\section{Introduction\label{sec:introduction}}
Stochastic cooling as a mechanism to shrink the particle phase space was proposed in 1972 \cite{sc:1972} and has been successfully implemented in a number of antiproton, proton, and heavy ion storage rings \cite{rhic_sc:2008, tevatron_sc:2011}. The cooling rate is limited by the number of particles in the bunch and the system bandwidth \cite{vandermeer:1985, marriner:2004}. In 1993 Mikhailichnko and Zolotorev proposed extending the stochastic cooling bandwidth to optical wavelengths by using undulators as pickup and kicker \cite{osc_prl:1993}. They suggested a gradient undulator as pickup so that the intensity of the radiation (and the momentum kick imparted in the kicker undulator) would be proportional to the displacement from the undulator axis. A year later, a transit-time method of OSC (TTOSC) was proposed by Zolotorev and Zholents \cite{ttosc:1994}. In the transit-time method, the intensity of the radiation is independent of the position in the
pickup undulator. Rather, the delay bypass is designed to couple betatron amplitude and momentum offset to the arrival time of the particle in the kicker undulator. The relative delay between radiation and particle is adjusted so that the momentum kick reduces emittance and momentum spread. Since then the dynamics of optical stochastic cooling have been explored theoretically and numerically in some detail, and experimental tests have been proposed \cite{iota:2013, bergan:2019}. Still, there is no experimental demonstration to date.

Although many efforts have been devoted to the study of OSC, there has not yet been a particle tracking study of the OSC process that accounts all major effects. Part of the reason is the simulation needs a realistic ring lattice with a reasonable bypass design, which requires dedicated resource and effort. In addition, simulating the incoherent heating effect from nearby particles can be challenging. The TTOSC theory is well developed \cite{ttosc:1994, lebedev:2014, sylee_nima:2004}. However, the synchrotron radiation (SR) damping and excitation is not included in the theory, which may be appropriate for hadron machines but not for lepton machines where strong SR take places. Thus, a realistic simulation of the OSC process including SR will be valuable to validate the bypass design and demonstrate OSC.

 \begin{figure}[b]
   \centering
   \includegraphics*[width=0.8\columnwidth]{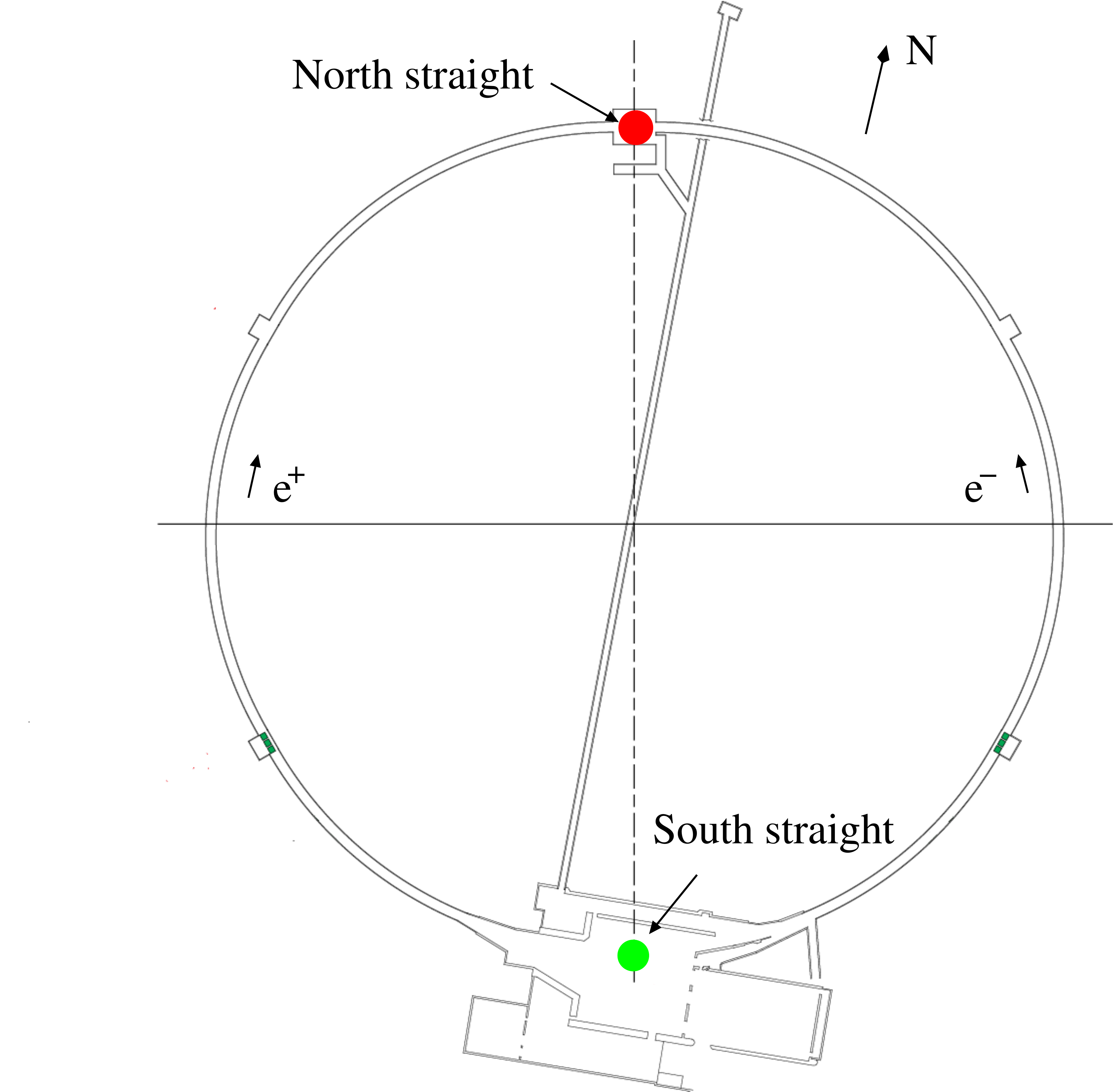}
   \caption{CESR layout showing two long straight sections.}
   \label{fig:cesrlayout}
\end{figure}

The Cornell Electron Storage Ring (CESR) built on the Cornell University campus stores counter-rotating beams of electrons and positrons and have operated as a collider for high-energy physics program for many decades. Currently, CESR serves as a synchrotron light source for x-ray users as well as a test accelerator for studying beam physics including electron cloud, intrabeam scattering, ion instabilities, and wake fields \cite{cesrta:2015}. The storage ring operates with beam energies that range from 1\unit{GeV} to 6\unit{GeV}.  The independent power supplies for all quadrupoles and sextupoles allow for a continuum of lattice configurations.
In CESR, positrons circulate in the clockwise direction and electrons in the counter-clockwise direction as shown in Fig.~\ref{fig:cesrlayout}. There are two long straight sections which originally served as the north and south interaction regions. Recently, the southern arc was reconfigured to install more undulators so as to accommodate more x-ray beamlines \cite{chessu:2019}. We plan to modify the north arc of the ring with beam optics to enable an experimental demonstration of OSC. Two types of bypass design have been considered. One early design is to modify the straight section ($\sim$10\unit{m}) along the northernmost rim of the ring to build a dog-leg bypass, consisting of 4 dipoles and a center defocusing quadrupole. This style bypass is similar to that proposed at Fermilab in the Integrable Optics Test Accelerator (IOTA) \cite{iota:2016}. We designed a flexible bypass layout that is compatible with path length delay up to 5\unit{mm}.  The other option is an arc-bypass design in which the light path is along a chord that intercepts 30$^{o}$ of the ring arc \cite{bergan:2019, andorf:2020}. The path length delay of this bypass layout is significantly larger $\sim$20\unit{cm}, which opens up the possibility of multi-pass or staged amplification schemes \cite{andorf:2020}. 

In this paper, we discuss three distinct bypass lines with the dog-leg type layout that have been matched into a full ring CESR lattice. Multiparticle tracking simulations that include the TTOSC process (coherent cooling and incoherent heating) are developed, and then used to characterize each set of optics, and in particular cooling times and dependence on bunch parameters. The tracking simulation confirms observable cooling for a 1-GeV bunch of $10^{7}$ particles. With more particles in a bunch, the horizontal profile of the beam shows a non-Gaussian shape during the OSC process, which could provide a useful signature of the OSC dynamics. The OSC damping rate and equilibrium emittance extracted from our simulation results without including SR effect agree reasonably well with the theory. In addition, our simulations confirm the phase space segmentation behavior expected from theory in the absence of incoherent kicks.  

The paper is organized as follows: in Sec.~\ref{sec:theory}, we briefly review the theory of TTOSC. In Sec.~\ref{sec:lat}, a CESR lattice including the bypass line layout is discussed. In Sec.~\ref{sec:method}, we describe the tracking simulation and in particular the method to account for the incoherent (heating) as well as the coherent (cooling) resulting from the coupling of optical radiation to particles in the kicker undulator. Results of simulation for the distinct sets of bypass optics are described in Sec.~\ref{sec:result}. Conclusions are summarized in the last section.

\section{Theory background\label{sec:theory}}
The TTOSC theory can be found in Ref.~\cite{ttosc:1994, lebedev:2014, sylee_nima:2004}. Here we briefly summarize the principles and reproduce some of the major formulas relevant to our discussion. 

A stochastic cooling system consists of a pickup, an amplifier, and a kicker. For optical stochastic cooling, both the pickup and the kicker are undulators which radiate with on-axis wavelength in the optical range. Stochastic cooling is an intrabunch feedback system. Radiation emitted by a particle at the characteristic wavelength of the pickup undulator is amplified  so that it can provide a momentum kick to that same particle in the kicker undulator, with phase shift suitable to reduce the particle's betaron and momentum error. In view of the relatively large undulator parameter K and long period required to generate undulator radiation at optical wavelengths (800\unit{nm}) for $\sim$1\unit{GeV} electrons, we plan to use helical rather than planar undulators. The advantage of helical undulator is that it results in a higher energy kick to a particle than a planar device when both have the same peak field, and also the desired wavelength can be obtained with lower peak field than a planar device \cite{helical:2018}.

The momentum kick to the particle is due to the interaction with radiation that the same particle emitted in the pickup undulator. We refer to this self-interaction as the coherent kick. The cooling derives from the coherent kick. The particle also receives incoherent kicks from the radiation from nearby particles. The noise from the incoherent kicks comprises a heating term. The momentum kick each particle receives in the kicker undulator is represented as \cite{ttosc:1994}
\begin{equation}
(\frac{\delta P}{P})_{i} = - G \sin(\Delta \phi_{i}) - G \sum_{j \neq i}^{N_s} \sin(\Delta \phi_{i} + \psi_{ij} ) \textrm{.}
\label{eq:kicks}
\end{equation}
Here $N_{s}$ is the number of particles moving behind the test particle $i$ within a distance of $N_u\lambda$, $N_u$ is the number of undulator periods, $\psi_{ij}$ is the radiation phase difference between the particle $j$ and the test particle $i$, and $\Delta \phi_{i}$ is the phase shift relative to the reference particle which receives zero momentum kick
\begin{equation}
\Delta \phi_{i} = k \Delta s \textrm{,}
\label{eq:cohere}
\end{equation}
where $k=2\pi/\lambda$ is the wave number; $\lambda=[\lambda_u(1+K^2)]/2\gamma^2$ is the wavelength of the first harmonic radiation emitted by the particles from the pickup undulator (helical), $\gamma$ is the Lorentz factor, $\lambda_u$ is the undulator period, and $K$ is the undulator parameter. In Eq.~\ref{eq:cohere}, $\Delta s$ is the particle's longitudinal displacement from the pickup to the kicker relative to the reference particle which receives zero kick. In a linear approximation where $\Delta s$ is small so that sin(k$\Delta s$) $\approx$ k$\Delta s$, it is written as 
\begin{equation}
\Delta s = M_{51} x + M_{52} x' + M_{56} \frac{\Delta P}{P} \textrm{,}
\label{eq:lin}
\end{equation}
where $M_{5n}$ are the elements of $6\times6$ transfer matrix from the pickup to the kicker, $x$, $x'$, and $\Delta P/P$ are the particle horizontal coordinate, angle, and relative momentum deviation at the pickup center.

If at the kicker undulator, the dispersion $\eta_k$ or the dispersion derivative $\eta'_{k}$ are nonzero, the betatron amplitude of the particle will change due to the the momentum kick according to
\begin{equation}
\Delta x_i = - \eta_k (\frac{\delta P}{P})_i \textrm{,    } \Delta x'_i = - \eta'_k (\frac{\delta P}{P})_i \textrm{.} 
\label{eq:xx}
\end{equation}
The changes to the horizontal phase space coordinates reduce the horizontal betatron amplitude, and thus cool the beam. If there exists $xy$ coupling in the machine, the vertical emittance can be reduced as well. The damping rates ($\lambda_{x}$, $\lambda_{z}$) derived in the linear approximation using the above relationships \cite{lebedev:2014} are
\begin{align}
\lambda_x &= \frac{kG (M_{56} - \widetilde{M}_{56} ) }{2} \label{eq:lx} \\
\lambda_z &= \frac{kG \widetilde{M}_{56}}{2} \label{eq:lz} \textrm{,}
\end{align}
where $\widetilde{M}_{56} = M_{51} \eta_p + M_{52} \eta'_p + M_{56}$. $\eta_p$ and $\eta'_p$ are the dispersion and dispersion derivative at the pickup undulator. 

When the particle has large oscillation amplitude such that $k\Delta s$ is large, the momentum kick will be nonlinear. Accounting for this nonlinearity by averaging kicks over betatron and synchrotron oscillations, one obtains the cooling boundaries ($\epsilon_{xmax}$, $(\Delta P/P)_{max}$) and the cooling ranges  ($n_x$, $n_z$) \cite{lebedev:2014} as
\begin{align}
\epsilon _{xmax} &= \frac{\mu^2}{k^2(\beta_p M_{51}^2 - 2 \alpha_p M_{51} M_{52} + \gamma_p M_{52}^2)}  \label{eq:emax}\\
(\frac{\Delta P}{P})_{max} &= \frac {\mu} {k \widetilde{M}_{56} } \label{eq:sigEmax}\\
n_x &= \sqrt{\frac{\epsilon_{xmax}}{\epsilon_x}} \\
n_z &= (\frac{\Delta P}{P})_{max} / \sigma_p
\end{align}
where $\beta_p$, $\alpha_p$, and $\gamma_p$ are the Twiss parameters at the center of the pickup undulator, $\epsilon_x$ is the horizontal equilibrium emittance without OSC, $\sigma_p$ is the energy spread of beam in the design lattice. The horizontal and longitudinal cooling rates are comparable when $\mu=\mu_0\approx2.405$. If either the horizontal or the longitudinal cooling dominates the other, the corresponding $\mu$ will be $\mu_{1}\approx3.832$ \cite{rubin:2017, iota:2016}. The $\mu_0$ and $\mu_1$ are the first zeros of the zeroth and first Bessel functions ($J_0$ and $J_1$), respectively \cite{lebedev:2014}.

The $\epsilon_{xmax}$ and $(\Delta P/P)_{max}$ define the boundaries within which the particles can be cooled by the TTOSC process. For a Gaussian beam $n_x$ and $n_z$ describe to how many sigmas in phase-space amplitude a particle can be and still be cooled. Large cooling ranges are desirable as it makes it harder for a particle to escape from the cooling region and be heated during, for example, a large intra-beam scattering event. An immediate drawback for a larger cooling range is a reduction in the damping rate for a fixed kick amplitude.

\begin{figure}[t]
   \centering
   \includegraphics*[width=0.8\columnwidth]{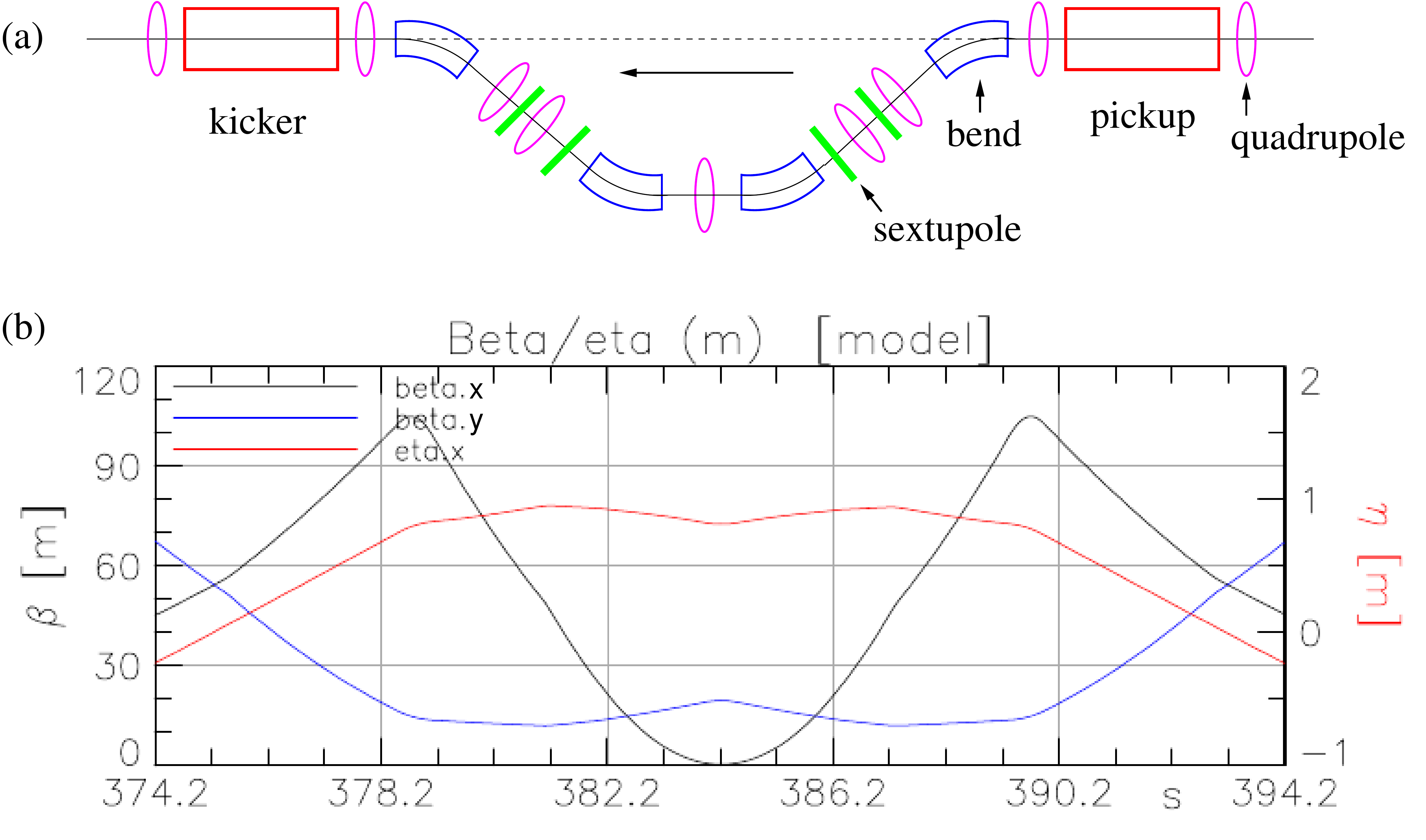}
   \caption{(a) Bypass layout matched to CESR. (b) The Twiss parameters $\beta$ and $\eta$ of the bypass optics. The horizontal and vertical betas are the black and blue curves respectively. The red curve shows the horizontal dispersion.  }
   \label{fig:bypass}
\end{figure}

In Ref \cite{sylee_nima:2004} Lee $\emph{et al}$ derived the damping decrements as well as the cooling dynamics equations when the incoherent heating is included. The horizontal damping decrement and cooling equation are listed as
\begin{align}
\alpha_x &= 2GkI_\perp e^{-u} - \frac{G^2N_sH}{2\epsilon_x} \label{eq:dampdec}\\
\frac{d\epsilon_x}{dt} &= -\frac{2GkI_\perp\epsilon_x}{T_0}e^{-u} + \frac{G^2N_s H}{2T_0} \label{eq:dampeq}\textrm{,}
\end{align}
while $u=\frac{1}{2}k^2[\beta_p M_{51}^2-2\alpha_p M_{51} M_{52}+\gamma_p M_{52}^2)\epsilon_x + \widetilde{M}_{56}^2\sigma_p^2]$ (the dispersion invariant at the kicker undulator), 
$H=\gamma_k\eta_k^2 + 2 \alpha_k \eta_k \eta'_k + \beta_k \eta'^{2}_k $ (curly H function at the kicker undulator), $T_0$ is the revolution period, and $I_\perp$ is a complicated term depending only on the bypass. The linear $G$ term in Eq.~\ref{eq:dampdec} describes coherent cooling while the quadratic term is from incoherent heating. The competition between coherent cooling and incoherent heating determines the OSC process. From Eq.~\ref{eq:dampdec} and \ref{eq:dampeq}, the optimum gain ($G_{opt}$) with maximum damping decrement and the equilibrium emittance ($\epsilon_{eq}$) can be found as
\begin{align}
G_{opt} &= \frac{2kI_{\perp}\epsilon_x e^{-u}}{N_s H} \label{eq:gopt} \\
\epsilon_{eq} &= \frac{GN_s H}{4kI_{\perp}e^{-u_{eq}}} \label{eq:e0} \textrm{.}
\end{align}
Here $\epsilon_{eq}$ is also included in $u_{eq}$, and can be obtained by numerically solving Eq.~\ref{eq:e0}. We will discuss these equations and compare them to the simulated results in detail in Sec.~\ref{sec:result}.

\section{Bypass lattice\label{sec:lat}}
The arrival in the kicker of the light emitted in the pickup undulator is delayed by the intervening optical elements (lenses, amplifiers, etc.) The electron beam must be similarly delayed so that it arrives in the kicker coincident with the radiation. The optics of the beam delay bypass also serve to couple betatron amplitude and momentum offset to path length. The delay length ultimately limits the complexity of the optical components in the path of the light. Two different types of bypass designs, namely the dog-leg chicane bypass and the arc bypass, have been studied at CESR. For the purpose of simulation, we focus on the dog-leg chicane bypass in this paper, although the experimental program at CESR is pursuing the arc-bypass which will be briefly discussed in Sec.\ref{sec:disc}.

The layout of the dog-leg bypass is shown in Fig.~\ref{fig:bypass}(a), which is designed to accommodate a delay of as much as 5.4\unit{mm} and beam energy over the range 300\unit{MeV} $< E_{beam} < $ 1.5\unit{GeV}. In this paper we discuss properties of bypass optics with 2\unit{mm} delay at 1\unit{GeV}. As shown in Fig.~\ref{fig:bypass}(a), the electrons pass the bypass line from the right to the left. The four quadrupoles near the pickup and kicker undulators are necessary to match the bypass line to the CESR ring lattice. The four bends create the required 2\unit{mm} delay for the electrons. The five quadrupoles between two outer bends are used to manipulate the optics to get adequate OSC cooling parameters. Four sextupoles are added to eliminate the nonlinearity of the flight path \cite{kafka:2015}. The Twiss parameters $\beta_x$, $\beta_y$, and dispersion $\eta_x$ of the 2mm-delay bypass line are shown in Fig.~\ref{fig:bypass}(b).

\begin{table}[b]
\centering
\caption{CESR machine parameters for OSC.}
\label{tab:osc}
\begin{tabular*}{\columnwidth}{@{\extracolsep{\fill}}lcr}
\hline
\hline
Beam Energy (GeV)             &$E_0$              & $1.0$\\
Circumference (m)             &$L$                &$768.438$\\
Transverse Damping time (s)  &$\tau_{x,y}$             & $0.5$\\
Longitudinal Damping time (s)  &$\tau_z$             & $0.25$\\
Momentum Compaction           &$\alpha_p$         & $0.006$\\
Nominal RF Voltage (MV)       &$V_\text{RF}$      & $0.6$\\
Synchrotron tune              &$Q_s$              & $0.027$\\
Horizontal tune               &$Q_x$              & $16.593$\\
Vertical Tune                 &$Q_y$              & $13.413$\\
Bunch length (mm)                 &$\sigma_z$              & $11.0$\\
Horizontal Emittance (nm\textperiodcentered rad)  &$\epsilon_x$     &$\sim 2.2$\\
Energy spread                  &$\sigma_p$              & $4.0\times10^{-4}$\\

\hline
\hline
\end{tabular*}
\end{table}

The parameters of the CESR lattice are summarized in Table~\ref{tab:osc}. The parameters of the helical pickup and kicker undulators are $N_u$=8, $\lambda_u$=32.5\unit{cm} and $K$=4.22 to yield the wavelength of the first harmonic light at 800\unit{nm}. The cooling boundaries with the 2-mm delay bypass line as well as other parameters are summarized in Table \ref{tab:para}. The cooling rates depend on the optical amplifier gain ($G$) but the ratio of the horizontal and longitudinal cooling rates is determined by the bypass optics according to Eq.~\ref{eq:lx} and \ref{eq:lz}. By design the horizontal damping dominates for the CESR bypass. The horizontal OSC cooling range depends on the equilibrium emittance of the storage ring which in turn depends on the number of particles in the bunch due to intrabeam scattering (IBS). For the design (zero current) emittance (2.2 nm$\cdot$rad), the horizontal cooling range $n_x$=3.3. Most of the particles will fall within the cooling boundary.

\begin{figure}
   \centering
   \includegraphics*[width=0.7\columnwidth]{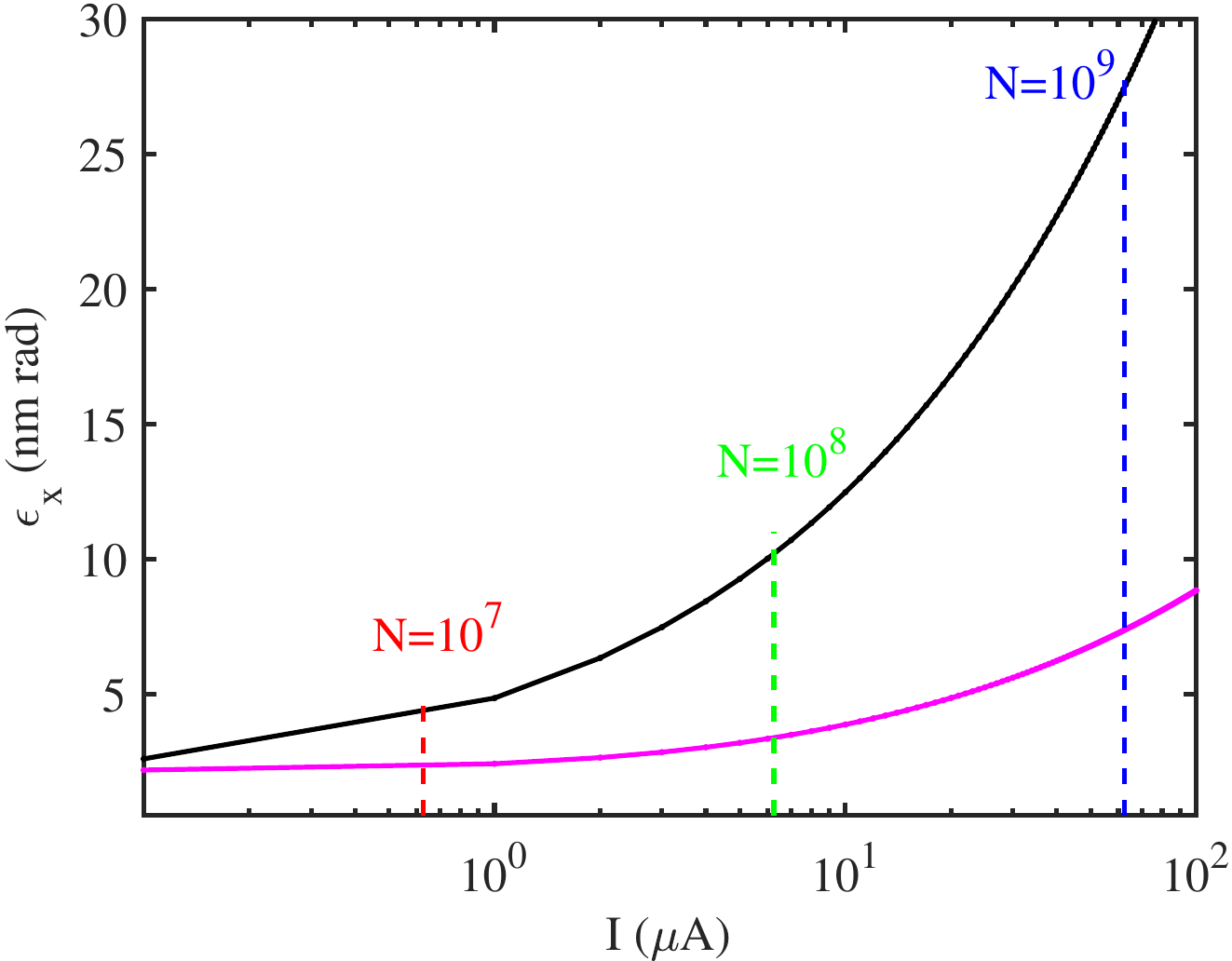}
   \caption{IBS-dominated horizontal emittance: no $xy$ coupling (black line) and $1\%$ $xy$ coupling (magenta line). The red, blue, and green dash lines indicate the CESR bunch currents with $10^7$,  $10^8$, and $10^9$ particles in the bunch, respectively.}
   \label{fig:ibs}
\end{figure}

The dependence of equilibrium emittance on bunch current is shown in Fig.~\ref{fig:ibs} \cite{prab-ibs}. With $10^7$ particles in a bunch ($I=0.625 \textrm{ } \mu A$), and assuming zero transverse coupling, the equilibrium emittance $\epsilon_x$ is 3.91 nm\textperiodcentered rad with an OSC cooling range of $n_x$=2.5. The emittance increases from 2.2 to 10.2 nm\textperiodcentered rad with $10^8$ particles in a bunch. For a bunch with $10^8$ particles, the ratio of horizontal cooling acceptance to equilibrium emittance drops to 1.5. The IBS emittance growth can be mitigated by introducing transverse coupling as indicated by the magenta line in Fig.~\ref{fig:ibs}.

The flexibility of the bypass and ring lattice will allow us to explore alternative sets of cooling parameters in both experiment and simulation.

\begin{table}[t]
\centering
\caption{Undulator and cooling Parameters}
\label{tab:para}
\begin{tabular*}{\columnwidth}{@{\extracolsep{\fill}}lcr}
\hline
\hline
$\lambda$ (nm)       & 800\\
$K_{u}$  &  4.22\\
$B_{u}$ (T) & 0.14\\
$\lambda_{u}$ (cm) & 32.5\\
$N_{u}$  & 8 \\
\hline
Bypass delay (mm) & 2 \\
$\epsilon_{max}$ (nm) & 24.1\\
$n_x$ & 3.3 \\
$(\Delta P/P)_{max}$ (\%) & 0.29 & \\
$n_z$ & 10 \\
$\lambda_x/\lambda_z$ & 30\\ 
\hline
\hline
\end{tabular*}
\end{table}

\section{Simulation methods\label{sec:method}}
The particle tracking is based on Bmad code library \cite{bmad:2006}. The bunch is modeled as a distribution of 1000 macroparticles, which is sufficient to calculate the bunch properties (emittance and beam size) accurately. If SR damping and quantum excitation (stochastic emission of photons) are switched on, the distribution relaxes to the equilibrium emittance and momentum spread after tracking for a few damping times. If quantum excitation and damping are switched off, the initial volume of the phase space is preserved. On each turn all the particles' 6-dimensional coordinates ($x$, $x'$, $y$, $y'$, $z$, $z'$) at the pickup are recorded in order to construct the sigma matrix, from which the three normal mode emittances are calculated \cite{wolski:2006}. On passage through the kicker undulator, each macroparticle receives both coherent and incoherent kicks. 

\subsection{Coherent kicks}
The coherent kick (see Eq.~\ref{eq:kicks}) depends on the gain $G$ and the differential path length $\Delta s$. The path from the pickup to the kicker (center to center) is recorded at each turn, $\Delta s=z_k-z_p$, where $z_k$ and $z_p$ are the same-turn longitudinal coordinates of the particle at the middle of the kicker and pickup undulator, respectively. The phase shift is calculated based on Eq.~\ref{eq:cohere}. The coherent kick $\Delta p_{z\textrm{-}co}=-G\sin(k\Delta s)$ is applied to each particle on every turn.

\begin{figure}
   \centering
   \includegraphics*[width=1.0\columnwidth]{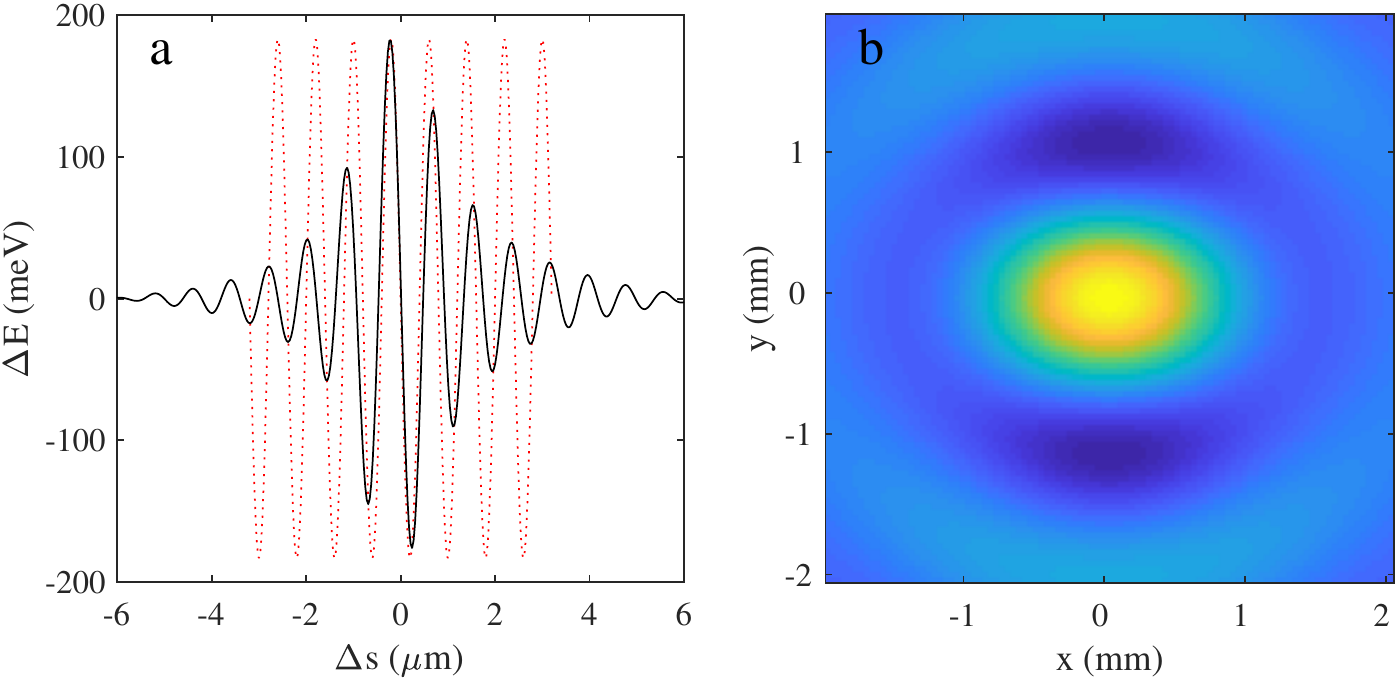}
   \caption{(a) Calculated energy change of an electron at the kicker as a function of the delay $\Delta s$ (black line). The red dotted line shows the energy change calculated from $-g\sin(k\Delta s)$ where $g=183$\unit{meV}. (b) Calculated transverse field $\mathcal{E}_{x}(x,y)$ of a single electron radiation from the pickup undulator. }
   \label{fig:coher}
\end{figure}

The above formula for the coherent kick assumes the undulators are long enough such that the particle delay is negligible compared to the finite pulse length emitted by a single electron, $\Delta s << N_u\lambda$. To exactly account for a finite pulse, a factor can be added: $\Delta p_{z\textrm{-}co}=-G(1-\frac{\Delta s}{N_u\lambda})\sin(k\Delta s)$ \cite{alex:2013}. 

To simulate the coherent kick more accurately, we calculate the energy change that an electron receives from interacting with its own radiation at the kicker undulator as a function of $\Delta s$ (Fig.~\ref{fig:coher}(a)) \cite{matt_nima:2018}. We then interpolate this energy-delay curve to determine the energy change ($\Delta E$) of a tracked particle based on its $\Delta s$ every turn. In addition, the energy change of an electron also depends on the transverse separation between itself and its radiation at the kicker as the calculated transverse field $\mathcal{E}_x(x,y)$ indicates in Fig.~\ref{fig:coher}(b). Since the energy change is proportional to the transverse field, the coherent momentum kick is scaled by the correction factor $\frac{\mathcal{E}_x(x,y)}{\mathcal{E}_x(0,0)}$: $\Delta p_{z\textrm{-}co}=\frac{\Delta E(\Delta s)}{E_0}\frac{\mathcal{E}_x(x,y)}{\mathcal{E}_x(0,0)}$. In simulation, we assume a single lens such that if a particle has a transverse position $x_p$ in the pickup, its light is focused to a transverse coordinate $-x_p$ in the kicker. Thus, in the kicker, if the particle has a coordinate $x_k$, it is separated from its radiation centroid an amount $x=x_k+x_p$. As the formula indicates, the separation depends on the bypass line. We find below in Sec.~\ref{sec:result} the RMS separations between the particle and its radiation are about $0.4$\unit{mm} horizontally and $0.03$\unit{mm} vertically with an initial particle-action of 2.2\unit{nm\textperiodcentered rad} and $1\%$ xy coupling. 

Without amplification, the maximum energy change of 183\unit{meV} shown in Fig.~\ref{fig:coher}(a) corresponds to a $G=1.83\times10^{-10}$ in Eq.~\ref{eq:kicks}. Thus a scale factor is applied to simulate a particular gain level while using the realistic energy-delay curve to calculate the energy kick.

\begin{figure}
   \centering
   \includegraphics*[width=0.87\columnwidth]{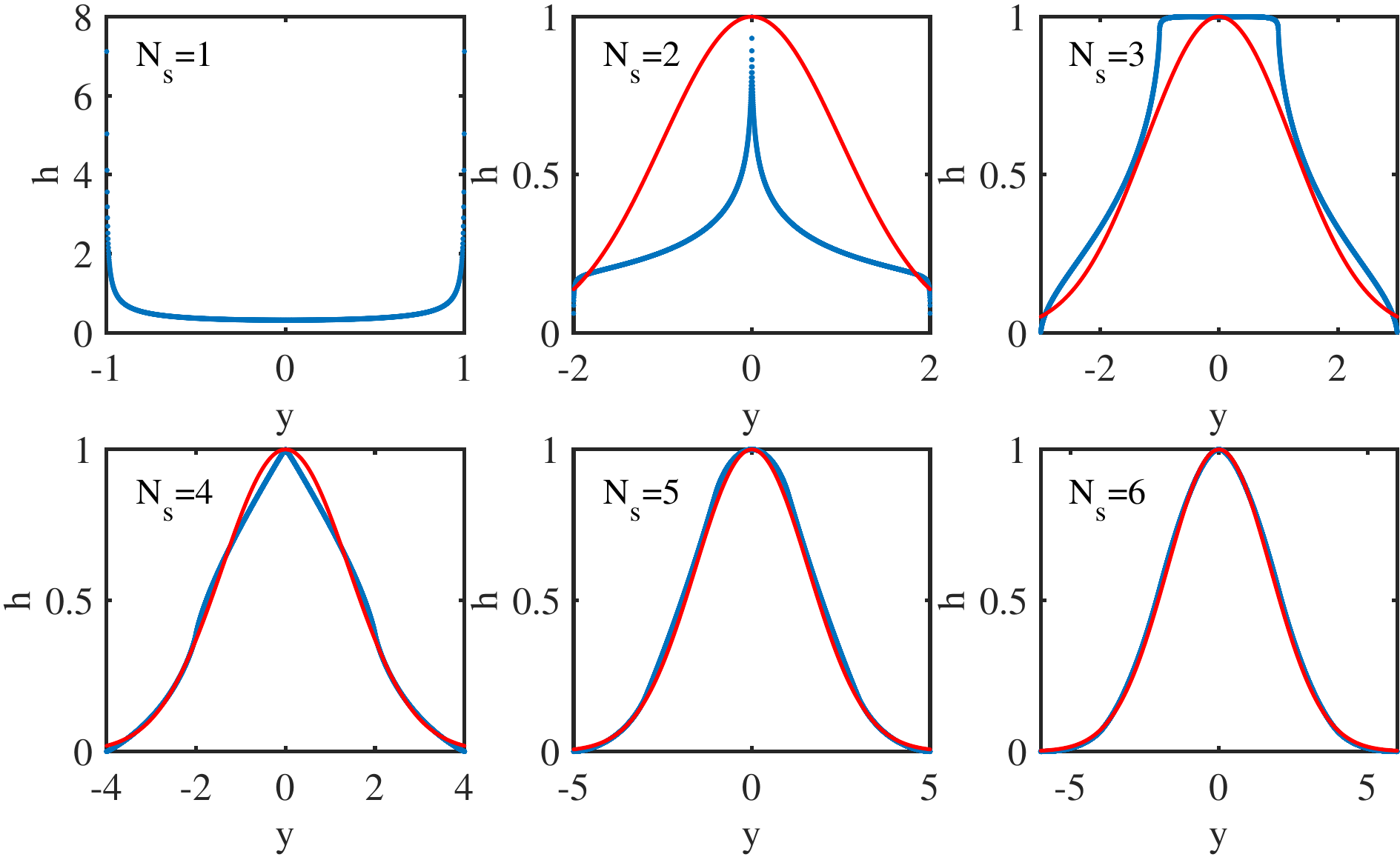}
   \caption{The normalized convolution function $h$ for $N_s$=1 to 6. The red lines are normalized Gaussian curves.}
   \label{fig:conv}
\end{figure}

\subsection{Incoherent kicks}
The second term on the right side of Eq.~\ref{eq:kicks} describes the incoherent contributions from the radiation of trailing particles. All particles trailing by distance less than $N_u\lambda$ behind the target particle will contribute to the noise. Conceptually, we simply could compute the differential path length for each of those trailing particles and add a kick with the appropriate phase. As long as there is a statistically representative number of particles within the range that contribute incoherent kicks, such a technique will be reliable. However, that turns out to be an impractically large number. For example, the simple approach would be to track a number of macroparticles equal to the actual number of particles in a bunch ($>10^7$), which is prohibitive. On the other end, if the bunch is comprised of 1000 macroparticles, then for an undulator with $N_u=8$  periods, and $\lambda=800$\unit{nm}, and bunch length $\sigma_z = 1$\unit{cm}, there is typically less than 1 trailing particle in the range of interaction with the leading particle, clearly insufficient to represent the incoherent heating.

It would be more convenient to apply a single kick to include the incoherent noise. We note that $\psi_{ij}=k(z_i-z_j)$ in Eq. \ref{eq:kicks} and $0 \leq z_i-z_j \leq N_u \lambda$, and therefore $\psi_{ij}$ is within the range $[0, 2N_u\pi]$. As long as $N_u\lambda<<\sigma_z$, we can assume that the particles are randomly distributed longitudinally within the slice $[z_i-N_u\lambda, z_i]$. Then $\psi_{ij}$ must be randomly distributed within $[0, 2N_u\pi]$. If $x$ is a random number within $[0, 2N_u\pi]$, the probability distribution function of $y=\sin(x)$ is
\begin{equation}
f(y)=\frac{1}{\pi \sqrt{1-y^2}} \textrm{, } y\subset [-1, 1] \textrm{. }
\end{equation}
The joint probability for a bunch, each with probability $f(y)$ is
$N_s$ convolutions of $f(y)$,  
\begin{equation}
h = f_1(y) * f_2(y) * \dots * f_{N_s}(y) \textrm{. }
\end{equation}
We find numerically, that for large $N_s$, the convoluted distribution approaches a Gaussian with a standard deviation proportional to $\sqrt{N_s}$. As shown in Fig.~\ref{fig:conv}, indeed for $N_s$ as few as 6, the probability distribution function $h$ is very nearly Gaussian. Evidently, the incoherent contribution (the second term of Eq.~\ref{eq:kicks}) can be simulated with a Gaussian function with width ($\sigma_{in}$) that scales with $N_s$, the number of particles in the slice. 

Consider the incoherent contribution from the particles in a central slice of a bunch of electrons. For a Gaussian distributed bunch with length $\sigma_z$, the number within the slice at $-N_u\lambda/2 < z <N_u\lambda/2$ is $N_s= N_{s\textrm{-}max}=PN$. Here $N$ is the total number of particles in the bunch and $P$ is the probability for a particle to be located within the range [-$N_u\lambda$/2, $N_u\lambda$/2]:
\begin{equation}
P=\int_{-\frac{N_u\lambda}{2}}^{\frac{N_u\lambda}{2}} \frac{1}{\sqrt{2\pi}\sigma_z} e^{-\frac{z^2}{2\sigma_z^2}}dz \textrm{.}
\end{equation}

\begin{figure}
   \centering
   \includegraphics*[width=0.65\columnwidth]{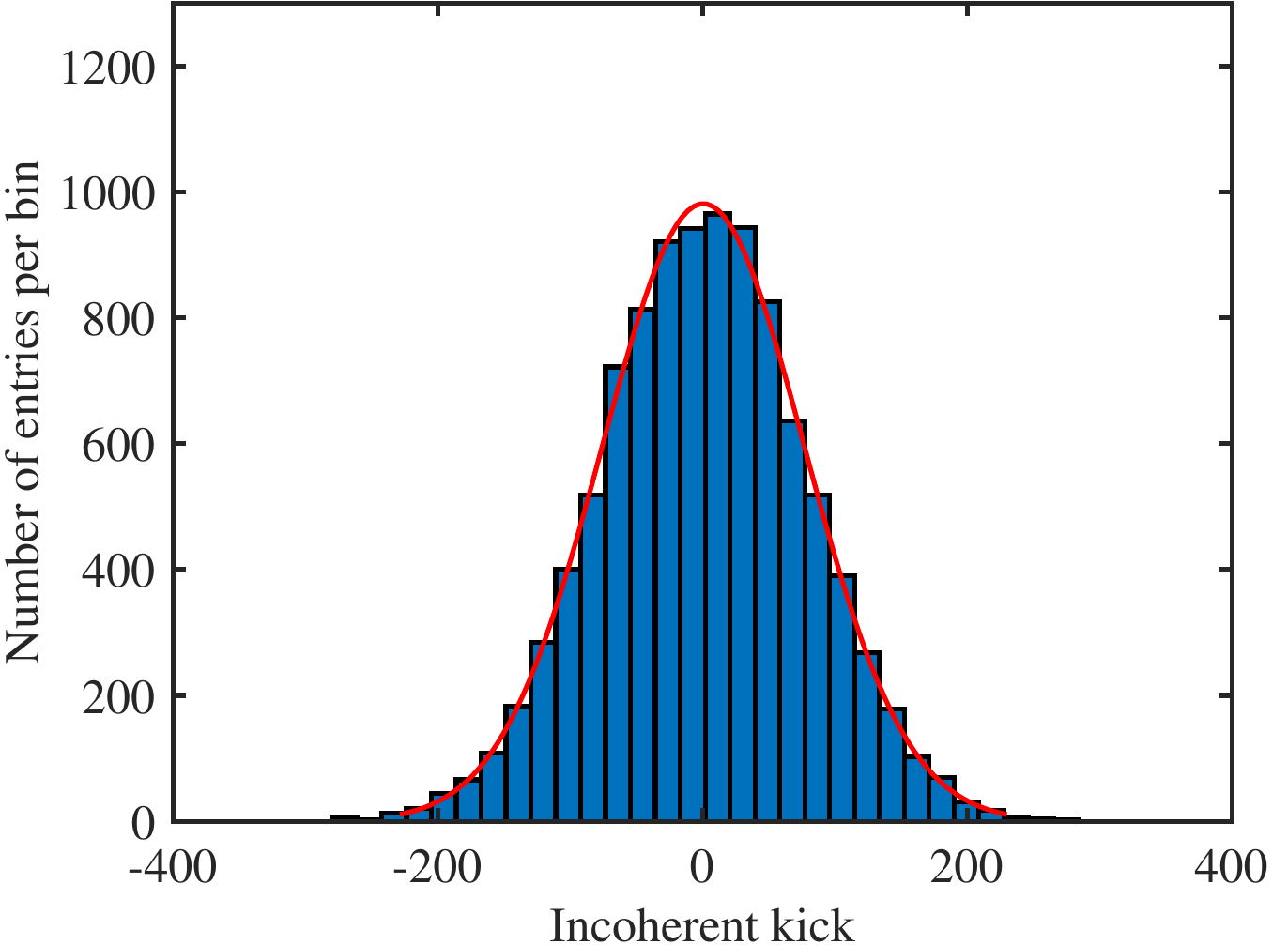}
   \caption{Histogram of $10^4$ simulated incoherent kicks with $N_s=11606$. The red line is the best Gaussian fit to the histogram.}
   \label{fig:inco}
\end{figure}

For $\sigma_z=11$ mm, $N_u=4$, $\lambda=0.8$ $\mu$m, and $N=10^8$, $P=1.1606\times10^{-4}$ so that $N_{s\textrm{-}max}$=11606. Kicks from each of $N_s$ randomly distributed particles are computed in a Monte Carlo simulation. Figure~\ref{fig:inco} shows the histogram of the resulting incoherent kicks. The distribution of the kicks is indeed Gaussian. The fitted standard deviation is $\sigma_{in\textrm{-}max}=76.24$. Furthermore, we find that, for a slice that is not at the bunch center ($z\neq0$) and if the bunch length $\sigma_z$ is different from $\sigma_0=11$\unit{mm} (e.g., shrinkage due to cooling), then the width of the distribution of incoherent kicks $\sigma_{in}$ is
\begin{align}
\sigma_{in}(z) &=\sigma_{in\textrm{-}max} \sqrt{\frac{N_s(z)}{N_{s\textrm{-}max}}} \label{eq:inco_gen}\\
               &=\sigma_{in\textrm{-}max} \sqrt{\frac{\sigma_0}{\sigma_z}} e^{-\frac{z^2}{4\sigma_z^2}} \label{eq:inco_gau}\textrm{,}
\end{align}
where $N_s(z)$ is the number of particles in the central slice of the bunch. The validity of Eq.~\ref{eq:inco_gau} requires that the particles are Gaussian distributed along the length of the bunch. If the longitudinal distribution is non-Gaussian, which could be generated by OSC as shown later in Sec.~\ref{sec:result}, Eq.~\ref{eq:inco_gen} will be used with $N_s(z)$ determined from the interpolation of the histogram of the $z$ distribution on every turn during the tracking simulation. 

Having determined the width $\sigma_{in}$, the incoherent kick is applied to each particle according to:
\begin{equation}
\Delta p_{z\textrm{-}in} = -G R \sigma_{in}(z) \textrm{,}
\end{equation}             
where $R$ corresponds to a random normal distribution. The total longitudinal kick $\Delta p_z = \Delta p_{z\textrm{-}co} + \Delta p_{z\textrm{-}in}$ is then applied to the particle at the exit of the kicker undulator on every turn to simulate the OSC cooling process.

As shown in Fig.~\ref{fig:coher}(a), the amplitude of the realistic energy change rapidly decreases below $10\%$ when $|\Delta s| >\frac{N_u\lambda}{2}$. This effect is neglected in the above incoherent kick calculation which could lead to an overestimate of the incoherent noise. To avoid the overestimate, we use $N_u$/2 as the undulator period in our incoherent kick estimation.

\section{Simulation results\label{sec:result}}

\begin{figure}
   \centering
   \includegraphics*[width=0.8\columnwidth]{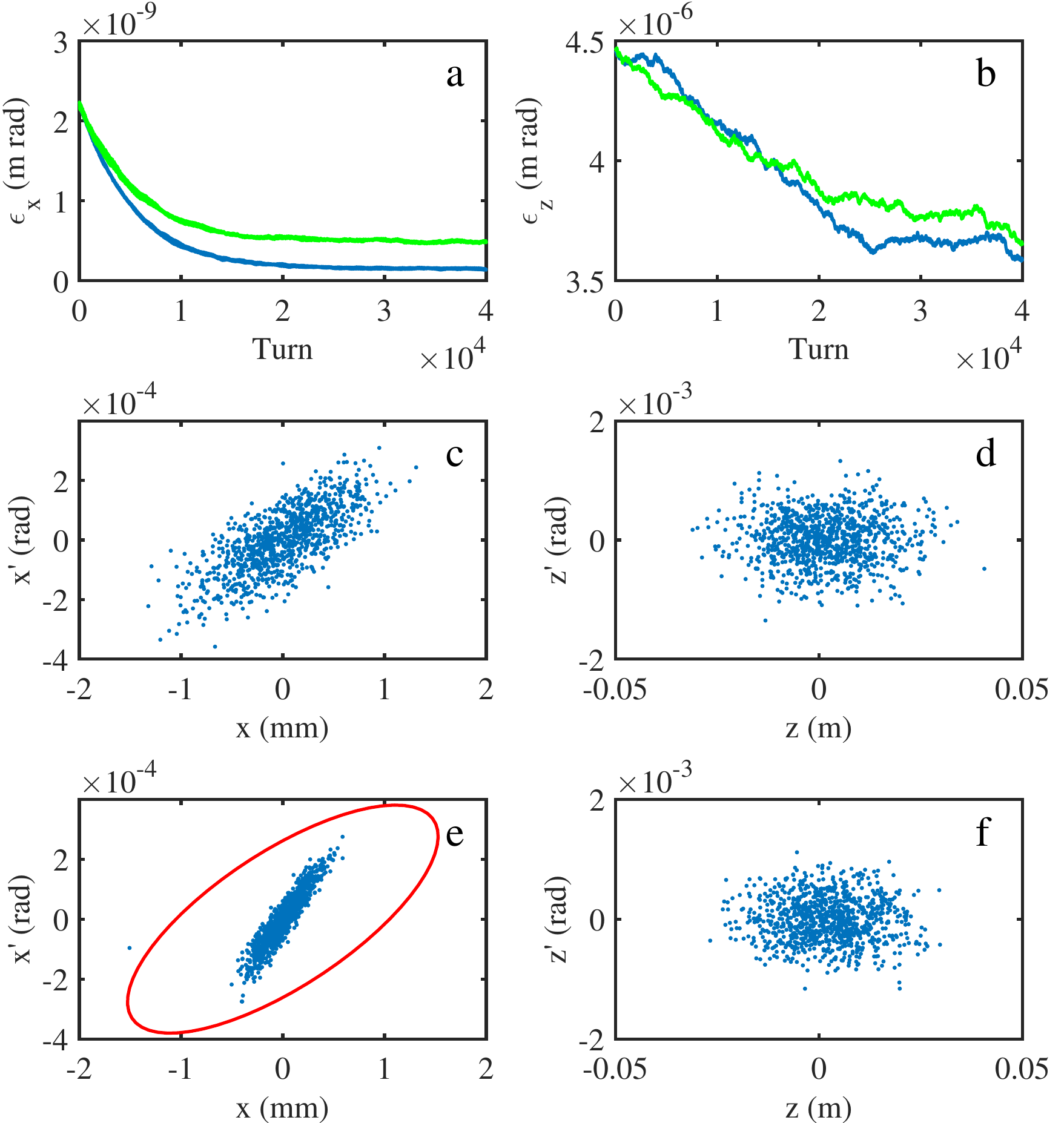}
   \caption{(a) Horizontal and (b) longitudinal emittances as functions of turns. The $xx'$ and $zz'$ phase spaces at turn 1 (c and d) and at turn 40000 (e and f). The red ellipse shows the cooling boundary $\epsilon_{xmax}$. The green lines are the results after adding incoherent kicks.}
   \label{fig:emit_des}
\end{figure}

\subsection{CESR lattice}
We demonstrate cooling in simulation by tracking a bunched distribution through the storage ring. The ring lattice parameters are summarized in Table~\ref{tab:osc} and the associated cooling parameters are in Table~\ref{tab:para}. We assume a bunch population of $10^7$ and 1\% emittance coupling so that equilibrium emittance is the single particle emittance (the contribution from IBS is negligible as indicated in Fig.~\ref{fig:ibs}). With SR damping and excitation switched on, the distribution relaxes to the single particle limit with horizontal emittance $\epsilon_x=2.2$ nm$\cdot$rad and longitudinal emittance is $\epsilon_z=4.5$ $\mu$m$\cdot$rad. In our initial simulation of the OSC process, only the coherent kick is applied. The gain $G$ was set at $10^{-8}$ for demonstration here. The effect of the coherent OSC kicks is to decrease both $\epsilon_x$ and $\epsilon_z$ in Fig.~\ref{fig:emit_des}(a) and (b), respectively. The horizontal and longitudinal phase space are shown in Fig.~\ref{fig:emit_des}(c) and (d) on turn one, and Fig.~\ref{fig:emit_des}(e) and (f) on turn 40000, and indeed indicate cooling of the distribution. The emittance shrinks to $\epsilon_x = 0.157$ \unit{nm\textperiodcentered rad} at turn 40000. 

\begin{figure}
   \centering
   \includegraphics*[width=0.8\columnwidth]{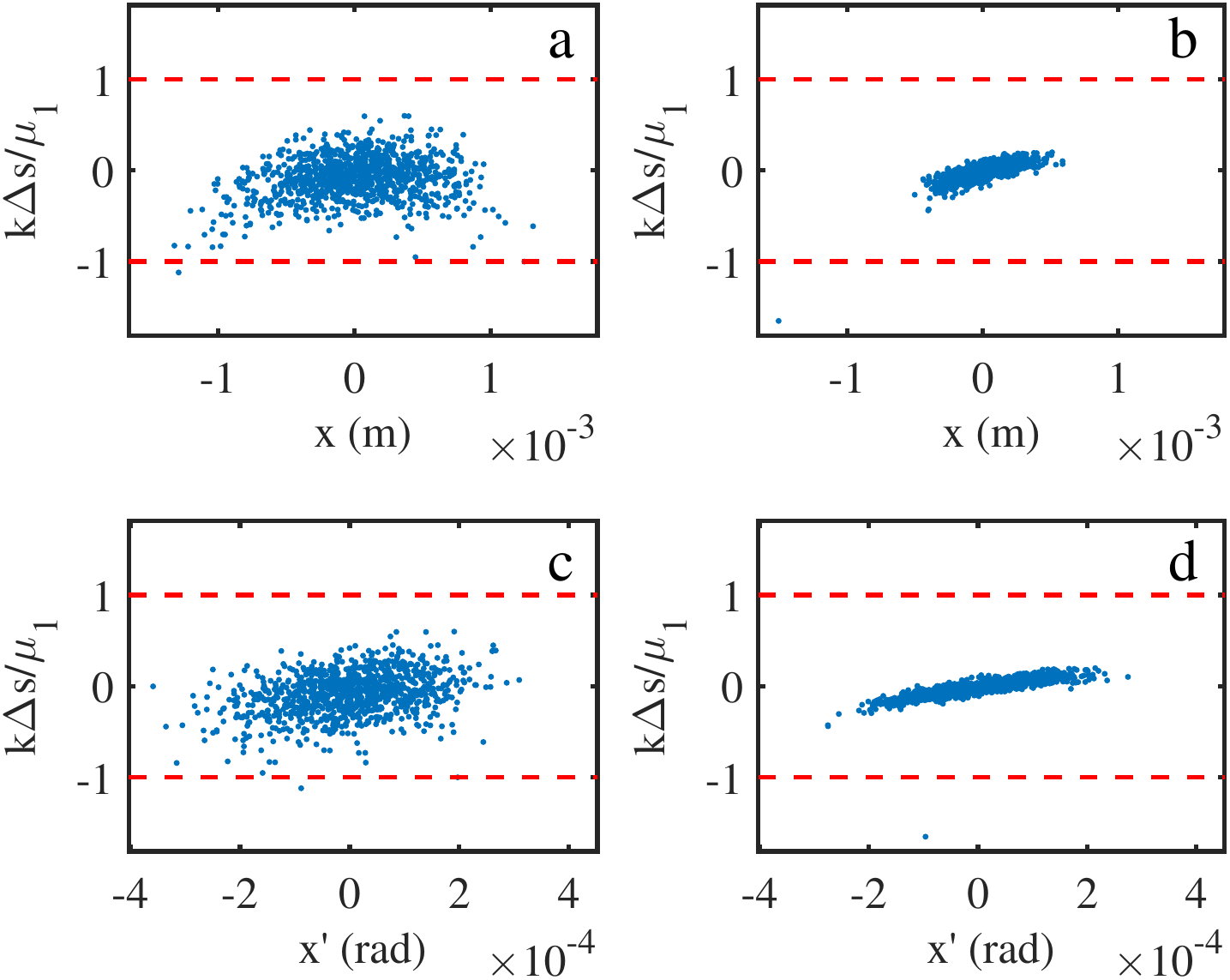}
   \caption{The normalized calculated $\Delta s$ of each particle as a function of its $x$ and $x'$ at turn 1 (a and c) and at turn 40000 (b and d). The red dashed lines show the cooling boundaries.}
   \label{fig:ds_des}
\end{figure}

Inspection of Fig.~\ref{fig:emit_des}(e), indicates that one particle out of the 1000 macro-particles is outside the cooling boundary (red ellipse in Fig.~\ref{fig:emit_des}(e)) and for that particle there is no reduction in betatron amplitude. Recall that the change in the electron's average path length $\Delta s$ through the bypass is a monotonically increasing function of the betatron amplitude in the pickup. If $k\Delta s > \mu_0$ ($\mu_0\approx$2.405, the first zero of $J_0$), the subsequent momentum kick will increase, rather than reduce that amplitude. The amplitude corresponding to $\Delta s = \mu_0/k$ defines the cooling boundary. The boundary that appears in simulation is consistent with theory \cite{lebedev:2014}. When the longitudinal cooling rate is small compared to the transverse rate ($\lambda_z \ll \lambda_x$) the transverse cooling boundary can also be represented as $k\Delta s/\mu_1 \leq 1$ ($\mu_1\approx$3.832). Figure~\ref{fig:ds_des} shows the delay ($\Delta s$) distribution for 1000 particles as a function of their phase space coordinates ($x$ and $x'$) on the first pass through the bypass ((a) and (c)) and then on turn 40000 ((b) and (d)). The bunch is cooled, and the phase offset $k\Delta s$ shrinks to near zero.

\begin{figure}
   \centering
   \includegraphics*[width=0.95\columnwidth]{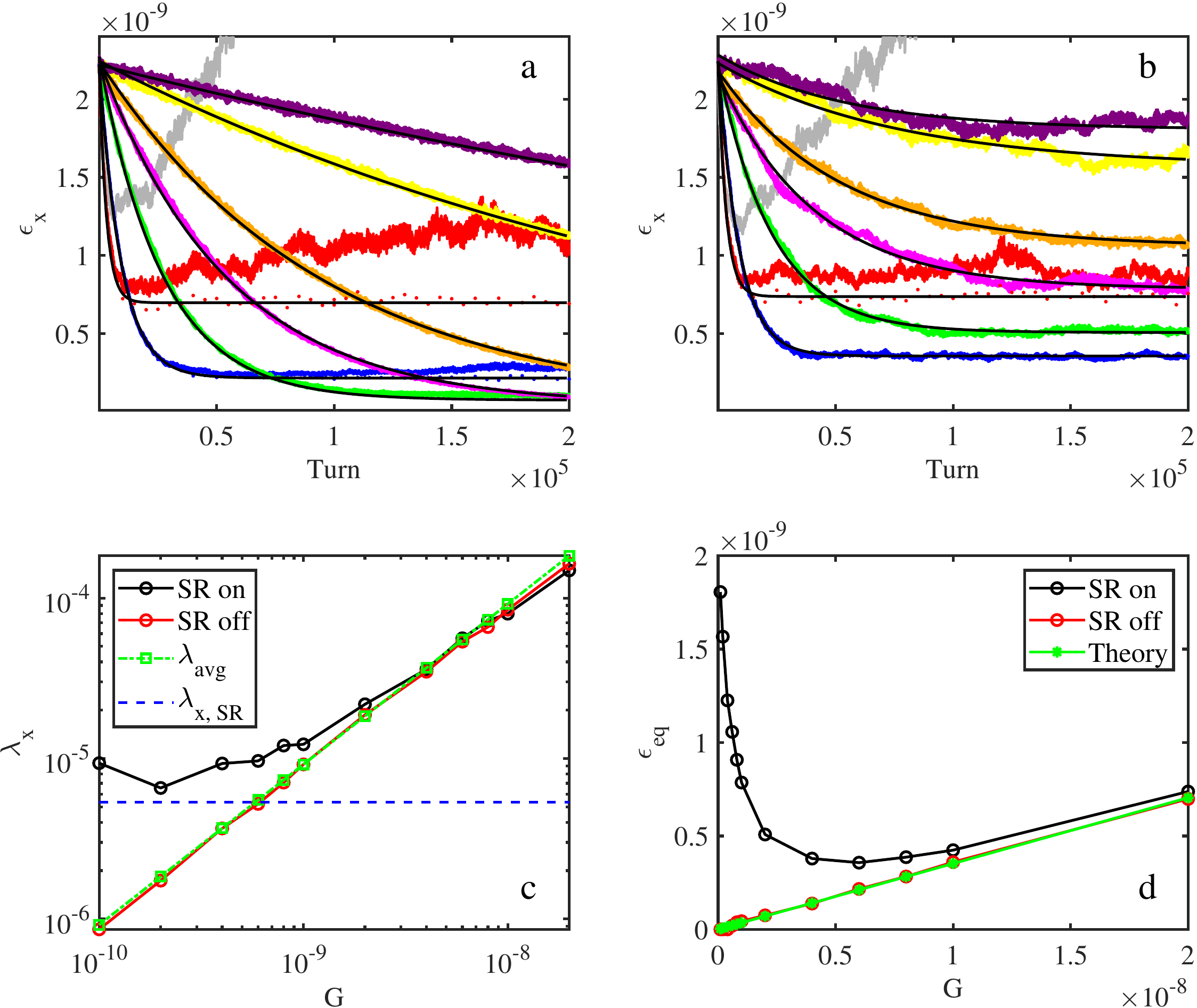}
   \caption{The horizontal emittances as a function of turn at various gain level: $3\times10^{-8}$ (grey), $2\times10^{-8}$ (red), $6\times10^{-9}$ (blue), $2\times10^{-9}$ (green), $1\times10^{-9}$ (magenta), $6\times10^{-10}$ (orange), $2\times10^{-10}$ (yellow), $1\times10^{-10}$ (purple), including both coherent and incoherent kicks, when SR damping and excitation is turn off (a) and on (b) in the simulation. The black solid lines are the best exponential fits. The extracted damping rates (c) and equilibrium emittances (d) from exponential fits to the data.}
   \label{fig:all_emit_2nm}
\end{figure}

The incoherent kicks are added to the simulation assuming $10^7$ particles in a bunch according to the same trend. Simulations assuming various levels of gain ($G$= $10^{-6}$, ... , $10^{-10}$) show that horizontal cooling is obtained when $10^{-10}\leq G \leq 2\times10^{-8}$ no matter whether SR damping and excitation is turned off or on (Fig.~\ref{fig:all_emit_2nm} (a) and (b)). When $G>3\times10^{-8}$, instead of cooling, the heating is observed. This is understandable because the coherent cooling depends linearly on $G$ while the incoherent heating is proportional to $G^2$ as Eq.~\ref{eq:dampdec} shows. When $G$ is large, the heating will dominate \cite{ttosc:1994, sylee_nima:2004}. From Eq.~\ref{eq:gopt}, we estimate the optimum gain is $G_{opt}\approx3\times10^{-8}$ while $u$=0.32, $I_\perp$=0.0032, $H$=2.64, and $N_s$=1000 are calculated from our bypass lattice and undulator parameters. As shown in Fig.~\ref{fig:all_emit_2nm} (a), the cooling is indeed observed at this optimum gain for the first 7000 turns and the horizontal emittance shrinks down to 1.25 nm$\cdot$rad. However, the beam starts to heat up drastically after 7000 turns. This is likely because the heating starts to dominate when $\epsilon_x$ decreases and $N_s$ increases to a certain level as Eq.~\ref{eq:dampdec} indicates. After 7000 turns, most particles are cooled and stay inside the core (within 1.25 nm$\cdot$rad) but the dominant incoherent heating gradually brings more and more particles outside the core and even outside the cooling boundary so that the emittance continues to increase. 

The OSC damping rates $\lambda_x$ and equilibrium emittances $\epsilon_{eq}$ obtained from exponential fits to the emittance cooling curves (Fig.~\ref{fig:all_emit_2nm} (a) and (b)) are plotted in Fig.~\ref{fig:all_emit_2nm} (c) and (d), respectively. The fitting function is $\epsilon_x=\epsilon_0 e^{-2\lambda_xt} + \epsilon_{eq}$, while $\epsilon_0$, $\lambda_x$, and $\epsilon_{eq}$ are the fitting parameters. As shown in Fig.~\ref{fig:all_emit_2nm}(a) and (b), the emittance curves at $G=2\times10^{-8}$ fluctuate a lot after 10000 turns. This is due to several particles outside the cooling boundary that are not cooled and that contribute noise to the emittance calculation. Therefore, we exclude them from emittance calculation to obtain new emittance curves (dotted symbols in Fig.~\ref{fig:all_emit_2nm}(a) and (b)) and then perform the exponential fits. For the results from simulation with SR damping and excitation switched off, the cooling rate increases with gain $G$, consistent with the theory \cite{lebedev:2014, sylee_nima:2004}. Note here Eq.~\ref{eq:lx} describes the damping for a single particle. For a bunch of particles, the average damping rate will be $\lambda_{avg}=\lambda_x e^{-u}$ \cite{sylee_nima:2004}. At higher gains, the damping rate deviates a little more from the theory, indicating more incoherent heating. The extracted equilibrium emittance also agrees well with the theory (Eq.~\ref{eq:e0}). For the results from simulation with SR damping and excitation turned on, both damping rate and equilibrium emittance deviate more from the theoretical calculation. This is understandable because SR damping and excitation are not included in the theory. At lower gains, the OSC process is weak and negligible compared to SR damping and excitation. When $G$ approaches zero, the total damping rate approaches the SR damping rate and the equilibrium emittance reaches the design zero current emittance. From Fig.~\ref{fig:all_emit_2nm} (d), there exists an optimum gain ($5\times10^{-9}$) at which the minimum emittance can be achieved from OSC. It is worth noting here that SR is strong in our lepton machine so that SR cannot be ignored.

\begin{figure}[t]
   \centering
   \includegraphics*[width=0.9\columnwidth]{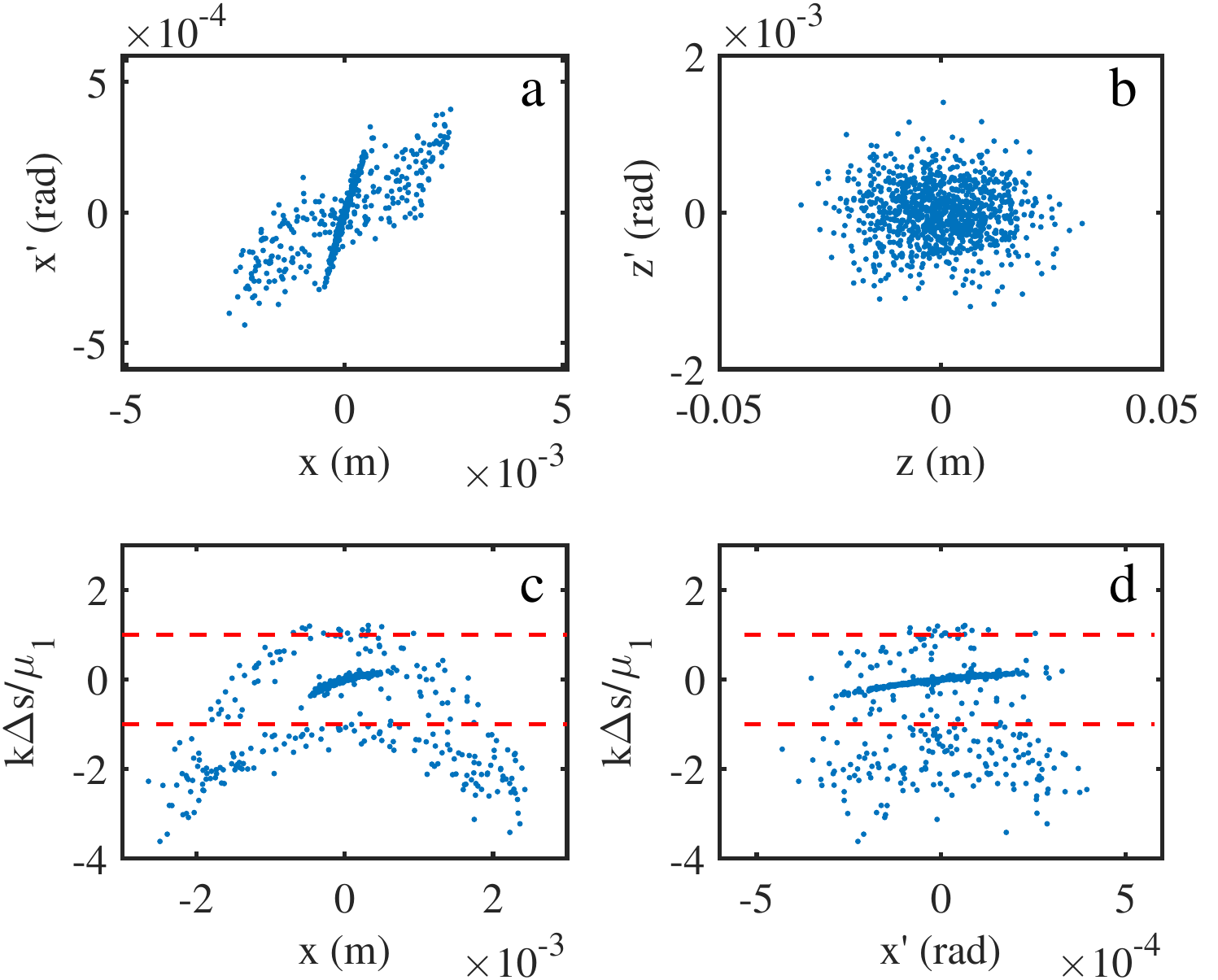}
   \caption{The $xx'$ (a) and $zz'$ (b) phase spaces after tracking the 1000 particles with initial emittance of 10 nm$\cdot$rad for 50000 turns. The normalized calculated $\Delta s$ of each particle as a function of its $x$ (c) and $x'$ (d) at turn 50000. The red dashed lines show the cooling boundaries.}
   \label{fig:ds_des_10nm}
\end{figure}

\begin{figure}
   \centering
   \includegraphics*[width=0.9\columnwidth]{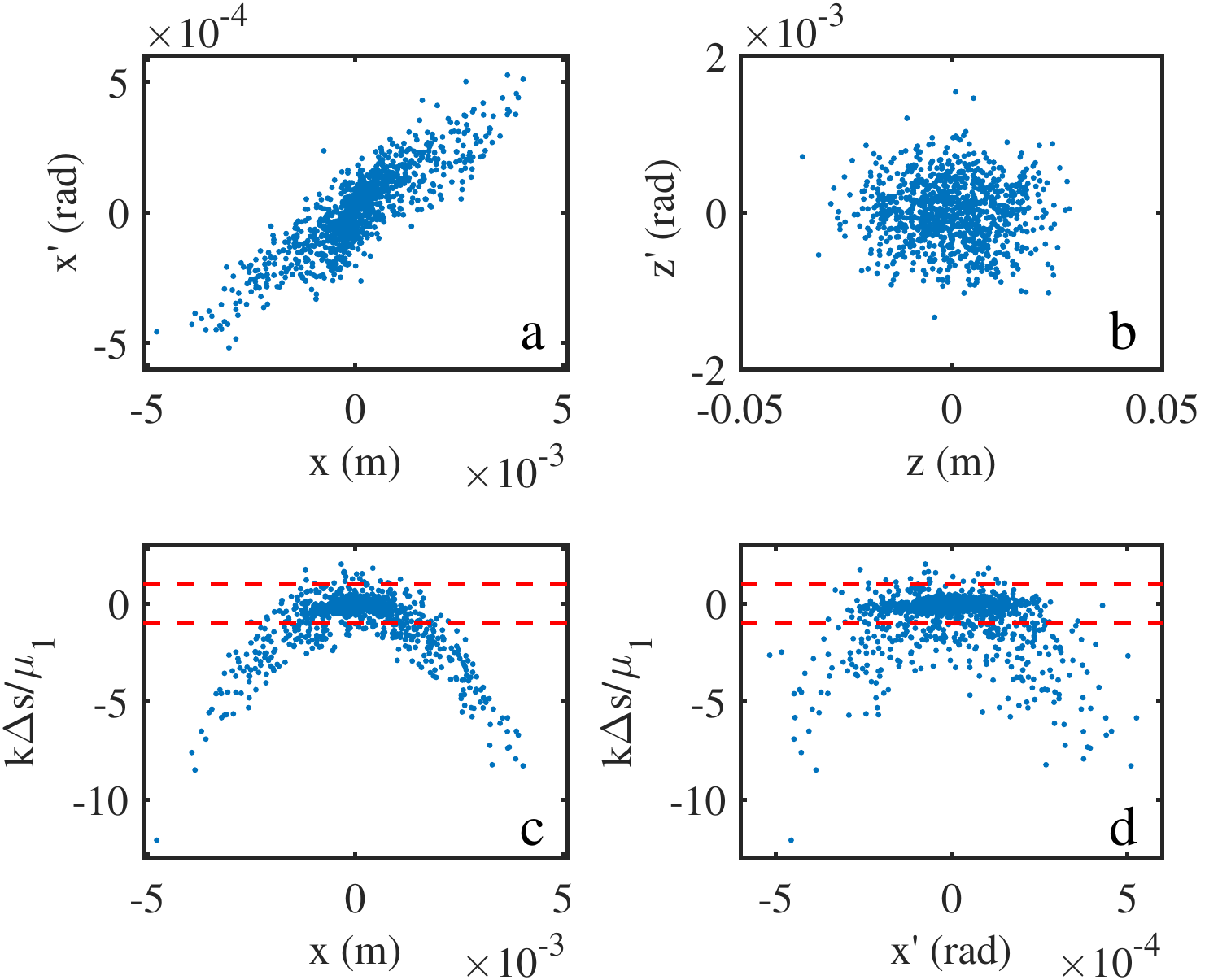}
   \caption{The $xx'$ (a) and $zz'$ (b) phase spaces after tracking the 1000 particles with initial emittance of 10 nm$\cdot$rad for 50000 turns including the incoherent kicks. The normalized calculated $\Delta s$ of each particle as a function of its $x$ (c) and $x'$ (d) at turn 50000. The red dashed lines show the cooling boundaries.}
   \label{fig:ds_des_10nm_inco}
\end{figure}

\subsection{Intra-beam scattering}

If the number of particles in a bunch is increased to $10^8$, the equilibrium horizontal emittance in the absence of stochastic cooling will grow to 10.2 nm$\cdot$rad due to intra-beam scattering, assuming no transverse coupling. In addition, the heating effect of the incoherent kicks increases with number of particles within each slice. Intra-beam scattering is not dynamically included in the simulation. In order to model the larger initial emittance due to IBS, the SR damping and excitation is switched off. The radiation damping will otherwise reduce (restore) the emittance to the zero current equilibrium. The phase space at turn 50000 with coherent but no incoherent kicks included is shown in Figure~\ref{fig:ds_des_10nm}(a) and (b). The horizontal cooling range for an initial 10 nm$\cdot$rad bunch is small $n_x=\left(\epsilon_{xmax}/\epsilon_x\right) =  1.5$, and many particles are outside the cooling aperture. 
As shown in Fig.~\ref{fig:ds_des_10nm}(a), the betatron amplitude is reduced only for those particles inside the cooling aperture. Particles outside the cooling aperture migrate to other locations in the phase space. This segmentation of the phase space is anticipated \cite{eoc:2008, ttosc:2012}. We will discuss this phenomenon in the next section in more detail. With the addition of the incoherent kicks, the segmentation is smeared as shown in Fig.~\ref{fig:ds_des_10nm_inco}.

\subsection{Bypass nonlinearity}

\begin{figure}
   \centering
   \includegraphics*[width=0.9\columnwidth]{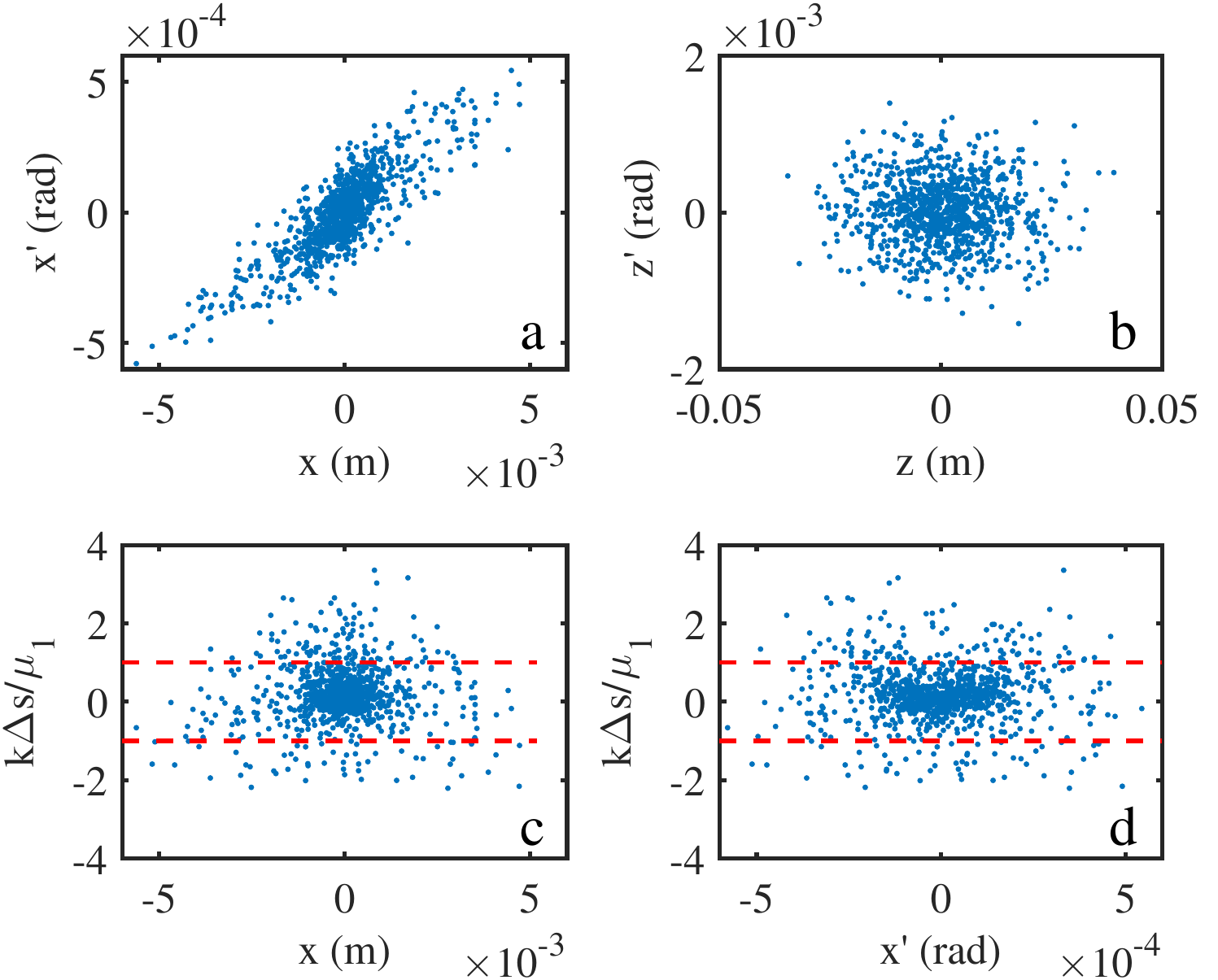}
   \caption{For the OSC bypass with sextupoles, the $xx'$ (a) and $zz'$ (b) phase spaces after tracking the 1000 particles with initial emittance of 10 nm$\cdot$rad for 50000 turns including the incoherent kicks after turning on bypass sextupoles. The normalized calculated $\Delta s$ of each particle as a function of its $x$ (c) and $x'$ (d) at turn 50000. The red dashed lines show the cooling boundaries.}
   \label{fig:ds_des_10nm_inco_sext}
\end{figure}

As Fig.~\ref{fig:ds_des_10nm}(c) and \ref{fig:ds_des_10nm_inco}(c) show, the phase delay at $k\Delta s$ has a nonlinear dependence on displacement $x$, $x'$ in the pickup. With the addition of four sextupoles into the bypass as indicated in Fig.~\ref{fig:bypass}(a), the nonlinearity of the particle path is mitigated \cite{lebedev:2015}. The sextupoles are turned off in Fig.~\ref{fig:ds_des_10nm_inco} as well as in Fig.~\ref{fig:emit_des} and \ref{fig:ds_des}, whereas Fig.~\ref{fig:ds_des_10nm_inco_sext}(c) and (d) show the tracking results after correcting the nonlinearity with the sextupoles. Linearity is very nearly restored and the number of particles within the cooling boundaries increases from $80\%$ to $90\%$ after correction.

\subsection{Phase space segmentation}
We described above the segmentation of the phase space if there are a significant number of particles outside of the cooling aperture (Fig.~\ref{fig:ds_des_10nm}). This behavior is due to the appearance of multiple cooling/heating boundaries in the phase space. For simplicity, if momentum cooling is considered exclusively and particles have no synchrotron motion (RF off), the boundaries are defined by multiple solutions to $\sin(k\Delta s) = 0$ (see  Eq.~\ref{eq:kicks}). When $\sin(k\Delta s)$ changes its sign, as $k\Delta s$ increases through $\pi$, the momentum kick likewise changes sign and will increase rather than decrease, the momentum offset. The boundaries in the longitudinal ($zz'$) phase space with only longitudinal cooling  are defined as
\begin{equation}
k M_{56} \frac{\Delta P}{P}  = 2 n \pi \textrm{,  }
\label{eq:phase_seg}
\end{equation}
where $n$ is an integer. When $0 \leq k\Delta s \leq \pi$, $\sin(k\Delta s)$ is positive, the particles will be cooled and evolve to an equilibrium where $k\Delta s=0$. When $\pi < k\Delta s \leq 2\pi$, $\sin(k\Delta s) < 0$,  the particle amplitudes are increased and attracted to the $n=1$ boundary where $k\Delta s=2\pi$. This behavior is discussed in Ref \cite{eoc:2008, ttosc:2012}.

When both the horizontal and longitudinal cooling are included, the emittance decrements averaged over betatron and synchrotron motion can be found as \cite{lebedev:2014}
\begin{align}
<\Delta\epsilon_x> &\propto - G J_1(kA_x)J_0(kA_z) \textrm{,  } \label{eq:ex} \\
<\Delta\epsilon_z> &\propto - G J_1(kA_z)J_0(kA_x) \textrm{,  } \label{eq:ez}
\end{align}
where $A_x$ and $A_z$ are defined as
\begin{align}
A_x &= \sqrt{\epsilon_x(\beta_xM^2_{51}-2 \alpha_x M_{51} M_{52}+\gamma_x M^2_{52})} \textrm{,  }  \label{eq:ax}\\
A_z &= \widetilde{M_{56}}\frac{\Delta P}{P} \textrm{.  } \label{eq:az} 
\end{align}
Here $\epsilon_x$ is the particle's invariant betatron amplitude. For cooling in both the horizontal and longitudinal planes, $kA_x \leq \mu_0$ and $kA_z\leq \mu_0$ are both necessarily satisfied, to obtain the cooling boundaries ($\epsilon_{xmax}$, $(\Delta P/P)_{max}$) as in Eq.~\ref{eq:emax} and \ref{eq:sigEmax}. 

Since the locations of the boundaries depend on the details of the bypass line beam optics (Eq.~\ref{eq:ex} to \ref{eq:az}), we can explore segmentation in a ``real" simulation by constructing a CESR ring lattice that includes a bypasss that may be incompatible with measurable cooling. We show how multiple boundaries appear in the phase space for two distinct sets of bypass optics with the OSC parameters listed in Table.~\ref{tab:bypass}. Both sets of optics show horizontal and longitudinal cooling but with very different cooling rates. In one instance ('Optics-A') the longitudinal cooling rate is much greater than the horizontal ($\lambda_l>>\lambda_x$) and in the other ('Optics-B') horizontal dominates over longitudinal cooling. Consequently, the horizontal emittance acceptance is large ($\sim$ 70 nm$\cdot$rad) but the energy acceptance is very poor in Optics-A while in Optics-B, horizontal emittance acceptance ($\sim$ 6.8 nm$\cdot$rad) is poor but the energy acceptance is large. 

\begin{table}[b]
\centering
\caption{OSC parameters of the bypass in CESR lattices.}
\label{tab:bypass}
\begin{tabular*}{\columnwidth}{@{\extracolsep{\fill}}lcr}
\hline
\hline
Bypass line	         & Optics A           & Optics B\\
\hline
$M_{51}$             &$-7.48\times10^{-4}$          &$3.17\times10^{-4}$\\
$M_{52}$ (m)            &$-5.81\times10^{-3}$             & $-1.38\times10^{-2}$\\
$M_{56}$ (m)            &$9.81\times10^{-3}$             & $3.59\times10^{-3}$\\
$\widetilde{M_{56}}$ (m)            &$9.56\times10^{-3}$             & $1.05\times10^{-4}$\\
$\epsilon_{xmax}$ (nm\textperiodcentered rad)          &$69.6$                & $6.77$\\
$(\Delta P/P)_{max}$          &$6.38\times10^{-5}$              & $2.91\times10^{-3}$\\
$\lambda_x/\lambda_z$           &$1/28.8$              & $33.1$\\
$E$ (GeV)            &$0.5$    & $1.0$ \\
$\epsilon_{x0}$ (nm\textperiodcentered rad)          &$5.0$                & $22.0$\\
$\sigma_E$            &$2.92\times10^{-4}$             & $4.07\times10^{-4}$\\
\hline
\hline
\end{tabular*}
\end{table}

\begin{figure}
   \centering
   \includegraphics*[width=0.9\columnwidth]{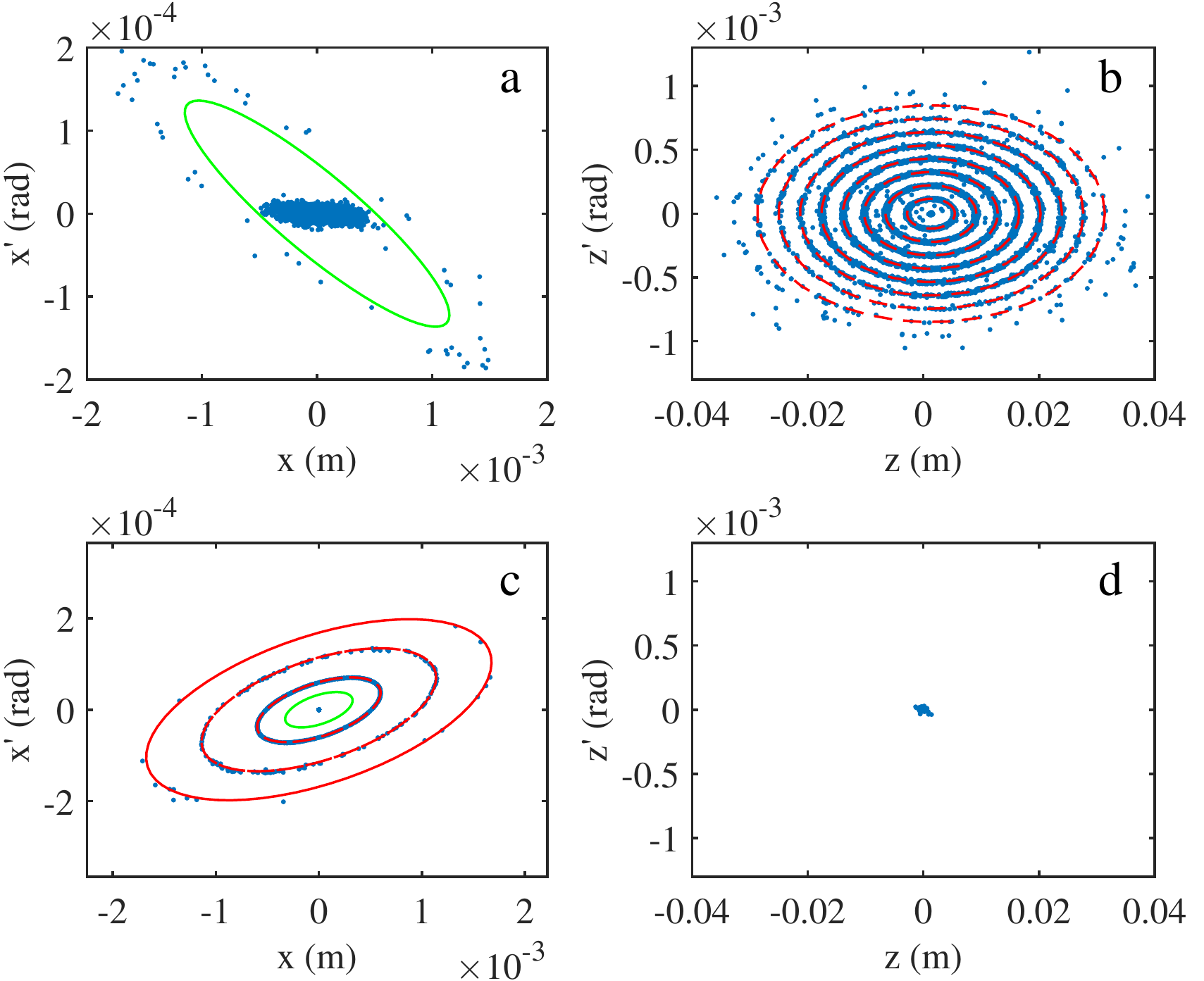}
   \caption{The $xx'$ and $zz'$ phase spaces for two lattices  including only coherent kicks and with $G=10^{-6}$ in the tracking simulation. Optics A: (a) and (b) $10^4$ particles at turn $10^5$; Optics B: (c) and (d) $10^3$ particles at turn $5\times10^4$. Dashed lines are the predictions see text for details. The green lines are the cooling boundaries.}
   \label{fig:phase_seg_zzp}
\end{figure}

\begin{figure}
   \centering
   \includegraphics*[width=0.9\columnwidth]{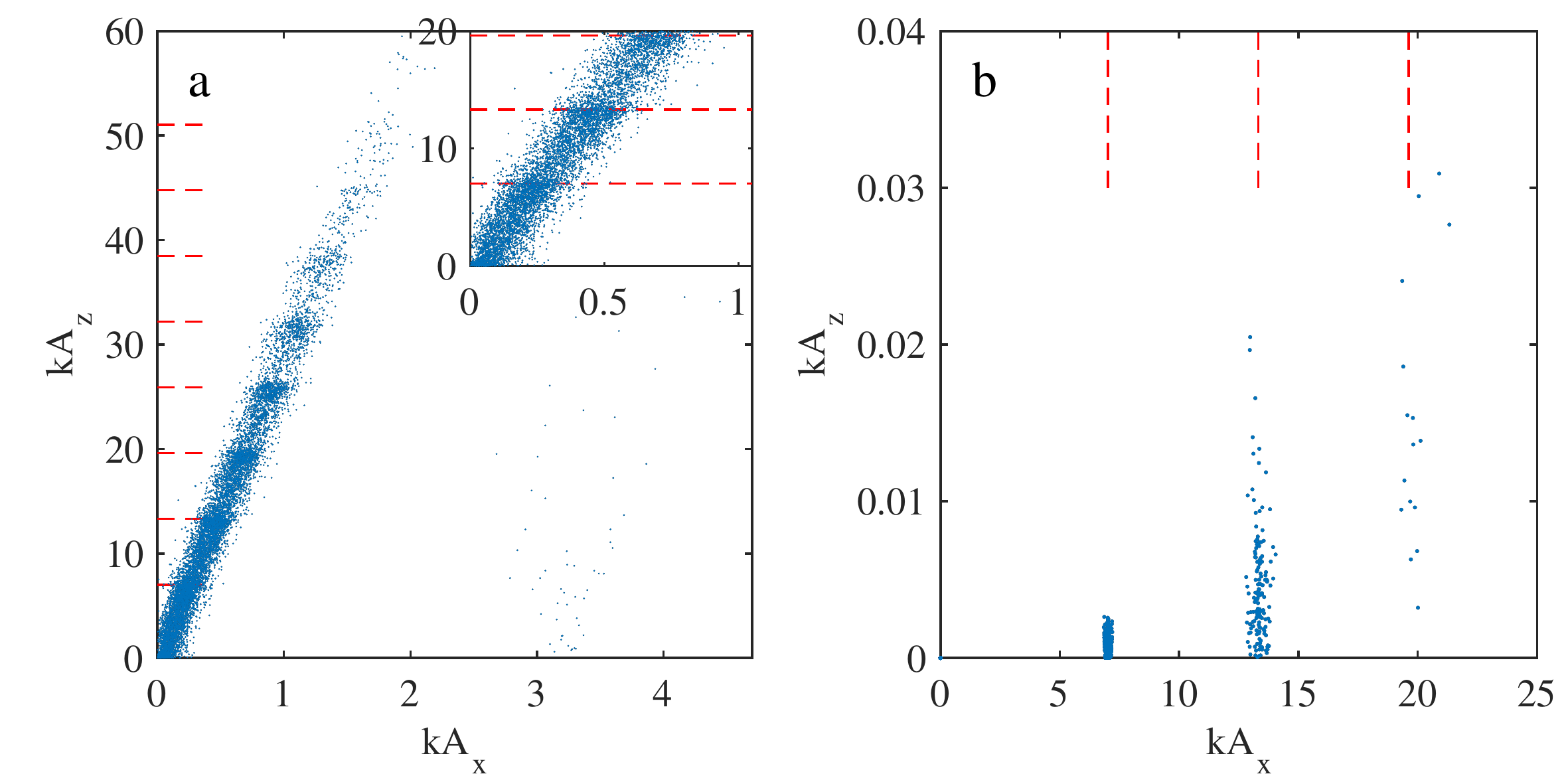}
   \caption{ $kA_z$ vs $kA_x$ of the particles in (a) Optics A and (b) Optics B shown in Fig.~\ref{fig:phase_seg_zzp}. The red dash lines indicates the $i$th zeros of the first Bessel function $J_1$ ($i=2, 4, 6, ...$): 7.016, 13.324, 19.616, ... . }
   \label{fig:phase_seg_axaz}
\end{figure}

Refering to Table~\ref{tab:bypass}, in the Optics-A bypass line $M_{51}$ and $M_{52}$ are small, and $M_{56}\approx\widetilde{M}_{56}$ so that the $\Delta s$ depends almost exclusively on the longitudinal synchrotron motion. In this case, when $kA_x\leq\mu_0$ and $J_{0}(kAx)$ is positive, the longitudinal cooling/heating will be determined by the sign of $J_1(kA_z)$. Thus, the cooling boundaries are determined by $J_1(kA_z)=0$, which defines the attraction rings in the longitudinal ($zz'$) phase space. That is $k\widetilde{M_{56}}\frac{\Delta P}{P}=\mu_{2n}$, where $\mu_{2n}$ is the $2n$ zeros of the first Bessel function. In the tracking simulation, the initial horizontal emittance of the bunch ($10^{4}$ particles) was set to $\epsilon_{x0}=5$ nm$\cdot$rad and the initial longitudinal distribution was set with the design values of bunch length and energy spread ($\sigma_z=10$ mm, $\sigma_E=2.92\times10^{-4}$) as determined by radiation excitation. In order to see the phase segmentation effect more clearly, we tracked the bunch for $10^{5}$ turns including only coherent kicks and with a high gain $G=10^{-6}$. Figure~\ref{fig:phase_seg_zzp}(a) and (b) show the $xx'$ and $zz'$ phase space at turn $10^{5}$. In the $zz'$ phase space (Fig.\ref{fig:phase_seg_zzp}(b)), the particles are indeed attracted to the fixed rings, in good agreement with the prediction ($k\widetilde{M_{56}}\frac{\Delta P}{P}=\mu_{2n}$, the red dashed lines). Figure~\ref{fig:phase_seg_axaz}(a) shows the $kA_z$ vs $kA_x$ of all the particles, displaying the particles indeed aggregate at $kA_z=\mu_{2n}$. In the $xx'$ phase space (Fig.~\ref{fig:phase_seg_zzp}(a)), the green line shows the horizontal cooling acceptance $\epsilon_{xmax}=69.6$ nm$\cdot$rad. The particles inside the boundary are cooled towards zero while the particles outside are heated and attracted toward the $\epsilon_{xmax}(\mu_1/\mu_0)^2$ boundary. 

The tracking simulation in Optics B was started with an initial bunch distribution of horizontal emittance of $22$ nm$\cdot$rad and design bunch length of $\sigma_z=10$ mm and energy spread of $\sigma_E=4.07\times10^{-4}$. The tracking was done with 1000 particles for $5\times10^4$ turns. In the Optics-B bypass line, $\widetilde{M}_{56}$ is small so that the $kA_z$ is small and the cooling/heating is determined by the sign of $J_1(kA_x)$. In order to observe the particle segregation effect, the gain level was set high ($G=10^{-6}$) and the incoherent kicks were not included. Figure~\ref{fig:phase_seg_zzp}(c) and (d) show the particle distributions in the $xx'$ and $zz'$ phase space respectively at turn $5\times10^{4}$. In the $zz'$ space, all the particles are inside the cooling range ($(\Delta P/P)_{max}=2.91\times10^{-3}$) so that they all migrate towards zero. In the $xx'$ space, the particles are attracted to the iso surfaces of emittance $\epsilon_{xn}$, satisfying $J_{1}(kA_x)=0$. Thus, $\epsilon_{xn}=\mu_{2n}^2/k^2/(M^2_{51}\beta_x+M^2_{52}\gamma_x-2M_{51}M_{52}\alpha_x)$. In Fig.~\ref{fig:phase_seg_zzp}(c), the dashed lines are plotted with emittances $\epsilon_{xn}$ using the Twiss parameters at the pickup undulator without the dispersion contribution. As we can see, the predicted rings reasonably match the simulation. It can also be seen clearly in Fig.~\ref{fig:phase_seg_axaz}(b) that the particles are attracted to $kA_x=\mu_{2n}$.

If the incoherent kicks are taken into account, the particle segmentation in the phase space is smeared and no sharp boundaries are observed. However, the microbunching structures could be evident if the gain is reduced to $G=10^{-7}$ and the bunch contains no more than $5\times 10^6$ particles.

\subsection{Discussion\label{sec:disc}}

In the proposed test of OSC at CESR, we expect to observe passive cooling. In this dog-leg style bypass layout, the maximum length of the pickup and kicker undulator is $\sim2.6$\unit{m}. With the undulator parameters $\lambda_u=32.5$\unit{cm} and $K=4.22$, the maximum energy gain of a single particle at the kicker undulator is estimated to be $\sim183$ meV \cite{lebedev:2014, helical:2018}, which corresponds to a $G=1.8\times10^{-10}$ at $1$\unit{GeV}. As Fig.~\ref{fig:all_emit_2nm} shows, a bunch with the emittance of $2.2$ nm$\cdot$rad is observed to be reduced to $1.6$ nm$\cdot$rad at $G=2\times10^{-10}$ with passive OSC including incoherent as well as coherent kicks. Observation of the expected emittance reduction using the synchrotron visible-light beam size monitor (vBSM) with a normal CCD camera will be difficult since the synchrotron radiation from $10^7$ particles in a bunch at $1$\unit{GeV} is much less than the intensity ($10^{10}$ particles at $5$\unit{GeV}) at which the vBSM normally operates \cite{suntao:2013}. Thus, a high sensitive camera will be extremely useful and necessary \cite{suntao:2017}. 

When the number of particles in a bunch increases, the beam emittance will increase due to IBS. The phase space is segmented as particles outside the cooling ranges are heated and attracted to the cooling boundaries. We observed the segmentation in simulation at the high gain level $G=10^{-6}$ and excluding incoherent kicks as shown in Fig.~\ref{fig:phase_seg_zzp}. With the addition of incoherent kicks, the phase space segregation is diluted, especially at lower gains. However, the emittance of the core part of the bunch that is within the cooling boundary is reduced. Therefore, in the proposed passive experiments, with many particles ($10^8 \sim 10^9$) in a bunch, observing distortion of the horizontal beam profile (Fig.~\ref{fig:ds_des_10nm}, \ref{fig:ds_des_10nm_inco}, \ref{fig:ds_des_10nm_inco_sext}) with direct imaging of transverse beam profile \cite{suntao:2013} will indicate OSC dynamics. 

As shown in Fig.~\ref{fig:ibs}, the horizontal IBS effect can be reduced and transferred to the vertical plane. Thus, with more particles in a bunch, increasing the $xy$ coupling will increase the cooling range in the horizontal so as to enhance the visibility of the OSC process. In addition, because of $xy$ coupling, the vertical emittance can be reduced along with the horizontal creating another observable signature of OSC.

Above we have discussed the simulation results on the dog-leg bypass layout. The same simulation principals apply to the arc-bypass layout. Compared to the dog-leg bypass, the arc-bypass design in CESR has two advantages: larger path length delay ($\sim20$\unit{cm}) and more space for the undulators ($\sim5.2$\unit{m}), leading to higher passive energy kick. However, the path lengths from the pickup to the kicker for both the particle and light are much longer, setting more strict stability requirement for the intervening dipoles. In Ref \cite{andorf:2020}, we have characterized the stability requirement of dipoles in this arc-bypass in detail. Similar particle tracking simulations were also performed to demonstrate the dependence of OSC damping rates on various levels of bend noise.

Finally, we note that our method to simulate the incoherent kicks can be applied to other transit-time cooling methods such as Coherent electron cooling (CeC) \cite{cec_prl:2009} and Microbunched electron cooling (MBEC) \cite{mbec_prl:2013}. Similarly, for these two cooling methods, the energy kick that an ion receives from its corresponding electron density spike (or bunched electrons) depends on the longitudinal distance $z$ between this ion and its corresponding electron density spike. Assuming the energy kick function is $f(z)$, which to result in cooing needs to be an odd function for z \cite{cec_func:2021}, besides this coherent kick, the ion will also receive incoherent kicks (shot noise) from random electron density spikes created by other ions: $\sum\limits_{i}^{N_s} f(z_i)$. Here $N_s$ represents the number of ions within a slice $[-z_0, z_0]$, that provide nonzero kicks to the ion. Similar to TTOSC, $\sum f(z_i)$ with random $z_i$ within the slice can be approximated by a Gaussian function. Then, tracking simulation including the incoherent kick for these two cooling methods will be treated similarly as TTOSC. There are other methods to apply random kick as the incoherent kick in CeC \cite{gang:2019} and MBEC \cite{mbec_prab:2020} processes as well. 

\section{Conclusion\label{sec:conc}}

We have developed tools to simulate realistic coherent kick as well as the incoherent noise in the TTOSC process. These simulation tools helped us understand the TTOSC concept and provided some guidance in the OSC lattice design. With both coherent and incoherent kicks included in the tracking simulation, cooling was observed in a bypass line at CESR at very low gain level ($G=2\times10^{-10}$, the passive cooling mode). In addition, the phase space segmentation was evident in two bypass lines with either small energy acceptance or emittance acceptance. These observations agree very well with the prediction from theory. 

\begin{acknowledgments}
The authors thank Mike Ehrlichman for designing early OSC lattices, Alexander Mikhailichenko, and Jim Shanks for valuable discussion and Dave Sagan for his assistance with Bmad; This research was supported by NSF award PHYS-1068662, PHYS-1416318, and PHYS-1549132, the Center for Bright Beams.
\end{acknowledgments}

\bibliography{prab_osc}

\end{document}